 \documentclass{aa}
\usepackage[utf8]{inputenc}

\usepackage{graphicx}
 \usepackage{txfonts}
 \usepackage{hyperref}
 \usepackage{booktabs}
 \usepackage{bm}
 \usepackage[normalem]{ulem}
 \usepackage[dvipsnames]{xcolor}
 \usepackage{nicefrac}
 \usepackage{xspace}

\usepackage{mathtools}
 \usepackage{hyperref}
 \hypersetup{
  colorlinks,
  citecolor=blue,
  linkcolor=blue,
  urlcolor=blue}
 \usepackage{amsmath}
 \usepackage{yhmath}
 \usepackage{multirow}
 \usepackage{array}
 \usepackage{geometry}
 \usepackage{footnote}
 \usepackage{tablefootnote}
 \usepackage{moresize}
\usepackage{placeins}

\newcommand{\Msun}{$\mathrm{M}_\odot$\xspace}
\newcommand{\Rsun}{$\mathrm{R}_\odot$}

\newcommand{\appropto}{\mathrel{\vcenter{
  \offinterlineskip\halign{\hfil$##$\cr
    \propto\cr\noalign{\kern2pt}\sim\cr\noalign{\kern-2pt}}}}}

\newcommand{\STELLA}{\texttt{STELLA}}

\begin{document}

\title{SN\,2023ixf -- an average-energy explosion \\with circumstellar medium and a precursor
 \thanks{The data computed and analysed for the current study are available via the link
 \href{https://zenodo.org/doi/10.5281/zenodo.13936548}{https://doi.org/10.5281/zenodo.13936548}.
}
}
\titlerunning{Average-energy SN\,2023ixf with CSM and precursor}


\author{
    Alexandra Kozyreva\inst{1},
    Andrea Caputo\inst{2,3},
    Petr Baklanov\inst{4,5,6}
    Alexey Mironov\inst{6}, \and
    Hans-Thomas~Janka\inst{7}
}
\authorrunning{A. Kozyreva et al.}
\institute{
    Heidelberger Institut f\"{u}r Theoretische Studien,
    Schloss-Wolfsbrunnenweg 35, D-69118 Heidelberg, Germany\\
    \email{sasha.kozyreva@gmail.com}
    \and
    Theoretical Physics Department, CERN, 1211 Geneva 23, Switzerland
    \and
    Dipartimento di Fisica, ``Sapienza'' Universit\`a di Roma \& Sezione INFN Roma1, Piazzale Aldo Moro 5, 00185, Roma, Italy
    \and
    National Research Center, Kurchatov Institute, pl. Kurchatova 1, Moscow 123182, Russia 
    \and
    Lebedev Physical Institute, Russian Academy of Sciences, 53 Leninsky Avenue, Moscow 119991, Russia
    \and
    M.V. Lomonosov Moscow State University, Sternberg Astronomical Institute, 119234, Moscow, Russia
    \and
    Max-Planck-Institut f\"{u}r Astrophysik, Karl-Schwarzschild-Str. 1, 85748 Garching bei M\"{u}nchen, Germany
    }

\date{Received; accepted }

\abstract{
The fortunate proximity of the Type II supernova (SN) 2023ixf allowed astronomers to follow its evolution from almost the moment of the collapse of the progenitor's core. SN\,2023ixf can be explained as an explosion of a massive star with an energy of $0.7\times10^{\,51}$~erg, however with a greatly reduced envelope mass, probably because of binary interaction. 
In our radiative-transfer simulations, the SN ejecta of 6~\Msun{} interact with circumstellar material (CSM) of (0.55--0.83)~\Msun{} extending to $10^{15}$~cm, which results in a light curve (LC) peak matching that of SN\,2023ixf. 
The origin of this required CSM might be gravity waves originating from convective shell burning, which could enhance wind-like mass-loss during the late stages of stellar evolution.
The steeply rising, low-luminosity flux during the first hours after observationally confirmed non-detection, however, cannot be explained by the collision of the energetic SN shock with the CSM. Instead, we considered it as a precursor that we could fit by the emission from (0.5--0.9)~\Msun{} of matter that was ejected with an energy of $\sim$$10^{49}$~erg a fraction of a day before the main shock of the SN explosion reached the surface of the progenitor.
The source of this energy injection into the outermost shell of the stellar envelope could also be dynamical processes related to the convective activity in the progenitor's interior or envelope. Alternatively, the early rise of the LC could point to the initial breakout of a highly non-spherical SN shock or of fast-moving, asymmetrically ejected matter that was swept out well ahead of the SN shock, potentially in a low-energy, nearly relativistic jet. We also discuss that pre-SN outbursts and LC precursors can be used to study or to constrain energy deposition in the outermost stellar layers by the decay of exotic particles, such as axions, which could be produced simultaneously with neutrinos in the newly formed, hot neutron star.
A careful analysis of the earliest few hours of the LCs of SNe can reveal elusive precursors and provide a unique window to the surface activity of massive stars during their core collapse. This can greatly improve our understanding of stellar physics and consequently also offer new tools for searching for exotic particles.
}


\maketitle

\section[Introduction]{Introduction} 
\label{sect:intro}

Supernovae (SNe) Type\,II, which have strong hydrogen lines in their spectra, are the most common stellar explosions within a limited volume \citep{2004ApJ...613..189D,2007MNRAS.377.1229M,2017PASP..129e4201S}. Stars that exhibit the properties of SNe\,IIP are massive stars that develop iron cores and, in turn, collapse and sweep away the entire envelope above. For many years, both observational and theoretical facilities have been attempting to solve the puzzles of this type of explosion. However, each new supernova (SN) brings us another piece of valuable information about the birth, childhood, youth, maturity, old age and reincarnation of stars. Similarly, SN\,2023ixf, discovered in May 2023 \citep{2023TNSTR..39....1I}, is one of the closest SNe in many years (at a redshift of $z = 0.000804$ and distance of 6.71\,Mpc). It opens a window for looking into the finer details of the evolution of stars and their explosions. Its proximity makes it a good laboratory for collecting and analysing data in detail.

Was SN\,2023ixf a regular SN\,IIP? Was its progenitor a massive star with an initial mass within the well-accepted range of $\sim$\,9 to 20~\Msun{}? Was the amount of energy released during the collapse of the core consistent with the current values predicted by first-principles multi-dimensional core collapse (CC) simulations? In our present study, we confront the SN parameters of SN\,2023ixf that have been reported in the literature. In the current literature, modelling of the light curves (LCs) and spectra of SN\,2023ixf suggests a range of values for the initial progenitor mass or ejecta mass, the radius of the progenitor, and the explosion energy. These values are listed in Table~\ref{table:references}. \citet{2023ApJ...954L..12T} derived their explosion parameters, such as high energy, while analysing the LC evolution over the first 19~days, therefore, there are no defined progenitor parameters such as an initial progenitor mass and the radius. Contrary to the results above, \citet{2024arXiv240807874H} inferred progenitors with larger initial masses (above 17~\Msun{}) which explodes with an energy of 0.7~foe and ejects less than 3~\Msun{} of hydrogen-rich matter. Simulations by \citet{2024arXiv240807874H} show relatively high fallback mass because of a lower explosion energy, in turn, resulted in a lower ejecta mass. The initial mass of a progenitor covers a wide range between 8 and 20 solar masses via progenitor identification, which is less certain than the mass estimates from direct LC or spectral modelling \citep{2023ApJ...952L..23K, 2023ApJ...955L..15N, 2023ApJ...953L..14P, 2023ApJ...957...64S,2024MNRAS.527.5366N, 2024MNRAS.534..271Q, 2024ApJ...968...27V, 2024SCPMA..6719514X}. Therefore, we do not include such data in the table. Parameters of the circumstellar matter (CSM) converge to a mass of $0.02-0.04$~\Msun{} and $0.4-0.85$~\Msun{}, and a radius of $(5-6)\,\times10^{14}$~cm. However, our main focus in this study is the value of the explosion energy estimated in the aforementioned studies. 

\begin{table}[!h]
\caption{Progenitor parameters of SN\,2023ixf in the literature.}
\resizebox{.5\textwidth}{!}{ \begin{tabular}{ c c c l } \hline
\Msun{}&\Rsun{}& foe\tablefootnote{foe\,$\equiv 10^{\,51}$~erg}   & References \\ \hline
10 & $\sim 500$ & 2--3 & \citet{2024PASJ...76.1050M}; \\
   &            &      & \citet{2024ApJ...975..132S}; \\ 
12 & $\sim 700$ & 1.2  & \parbox{5cm}{ \citet{2024A&A...681L..18B}; } \\
15 & 500--900   & 1--3 & \citet{2023ApJ...956L...5B}; \\
   &            &      & \citet{2023ApJ...954L..42J}; \\
   &            &      & \citet{2023ApJ...955L...8H}; \\
   &            &      & \parbox{5cm}{ \citet{2024A&A...683A.154M}. } \\
   &            & 2--5 & \citet{2023ApJ...954L..12T}; \\
16 &            &      & \citet{2025ApJ...978...36F}; \\  
17.5 -- 21.5 & $\sim 950$ & 0.48--0.71 & \citet{2024arXiv240807874H} \\ \hline
\end{tabular}  }  \label{table:references}   
\end{table}

The characteristic energy of SN explosions is roughly 1~foe. 
This is compatible with theoretical expectations that CCSNe of Type~II are caused by what has become known as the convectively supported neutrino-driven explosion mechanism \citep[e.g.,][for a selection of pioneering work and some more recent reviews]{Herant+1994,Burrows+1995,Janka+1996,Janka2012,Burrows2013,Mueller2016,Janka2017,Burrows+2021}. 
Spherically symmetric calculations of such neutrino-driven explosions using parametric neutrino ``engines'' \citep{2012ApJ...757...69U,2015ApJ...801...90P,2016ApJ...818..124E,2016MNRAS.460..742M,2016ApJ...821...38S,2019ApJ...870....1E} as well as three-dimensional (3D) neutrino-driven CC simulations show that explosion energies range between 0.05~foe and 1.4~foe for progenitors with initial masses between 9~\Msun{} and 20~\Msun{} \citep{2017MNRAS.472..491M, 2020MNRAS.496.2039S,2021ApJ...915...28B,2024ApJ...964L..16B,2024arXiv240113817J}. Observational estimates of the explosion energy are, in many approaches, based on fitting the observed spectra and LCs by means of radiation-hydrodynamics simulations and spectral synthesis, and are model-dependent. However, numerous studies converge to relatively similar numbers. For example, \citet{2022A&A...660A..41M} defined the range of explosion energies between 0.1~foe and 1.4~foe, with a median value of 0.6~foe. \citet{2003ApJ...582..905H} estimated energies for ``classical'' SNe\,IIP in a range between 0.6~foe and 5.5~foe. However, that study includes some unusual SNe, such as the bright SN\,1992am with an explosion energy of 5.5~foe and a mass of radioactive nickel $^{56}$Ni of 0.26~\Msun{}. This SN may not necessarily be powered by the neutrino-driven mechanism\footnote{\citet{2014A&A...565A..70K} showed that the LC of the bright SN\,1992am resembles the LC of a low-mass pair-instability SN, i.e. a thermonuclear explosion of a star with an initial mass of about 150~\Msun{}. It might therefore not necessarily be a neutrino-driven CCSN.}. Similarly, \citet{2015ApJ...806..225P} gave values for the explosion energies of 0.8--3~foe. Note however that \citet{2003ApJ...582..905H} and \citet{2015ApJ...806..225P} based their estimates on scaled relations that do not include radioactive nickel heating \citep{1993ApJ...414..712P}. Therefore they may have overestimated the true explosion energies. 
X-ray observations of SN remnants being less model dependent indicate characteristic CCSN energies in the range of $0.4-0.75$~foe \citep[see, e.g., ][]{1972ApJ...178..169C,2023MNRAS.521.5536K}.

Since SN\,2023ixf was a ``normal'' Type\,IIP SN, we expect that this SN exploded with an average energy, the value of which did not exceed the median of 0.6~foe. \citet{2024PASJ...76.1050M} applied an automatic procedure that provides a model with a progenitor of the lowest possible mass, because a lower ejecta mass is one of the ingredients to reduce the plateau duration. It might be that higher-energy CC explosions happen in nature, and there are observed SNe which exclusively require high explosion energies, such as stripped-envelope SNe and ``hypernovae'' \citep{1998Natur.395..672I, 2008Sci...321.1185M}. However, this is likely not the usual case for hydrogen-rich CCSNe, as otherwise the distribution of SNe\,IIP would extend up to claimed energy of 5~foe. However, a number of studies do not support such higher values. An automatic search based on a limited set of models may mistakenly result in a biased conclusion concerning the actual SN parameters. Similarly, a parameter estimate based on semi-analytic models and Monte-Carlo-Markov-Chain methods might not necessarily lead to a physically consistent scenario for SN explosions \citep{2020MNRAS.496.3725J}.

A reduction of the explosion energy produces a longer-lasting extended plateau, which is not compatible with the 85~day SN\,2023ixf plateau, which is shorter than the majority SNe\,IIP with an average plateau of 110~days \citep{2014ApJ...786...67A}. To compensate the increase of the length of the plateau, the hydrogen-rich ejecta mass has to be reduced accordingly. The lower energy also lowers the luminosity on the plateau, therefore, a larger radius of the progenitor is needed to obtain a luminosity on the same level.

The observations by \citet{2023ApJ...953L..16H} and \citet{2024Natur.627..754L} are unique among those obtained for SN\,2023ixf, because they show a clear flux excess during the first five hours of the SN. Although there was no pre-explosion outburst detected \citep{2023ApJ...957...28D,2024MNRAS.527.5366N}, we analyse the earlier data in the context of such an outburst, which might not necessarily have occurred months to years before the CC explosion but much shorter before it.

In the present study, we construct a progenitor structure suitable to explain the global observational properties of SN\,2023ixf. We introduce our progenitor configuration and a sample of CSM structures in Section~\ref{sect:method}. In Section~\ref{sect:res}, we describe how our modelled LCs compare to observed properties of SN\,2023ixf. We pay special attention to the first 5~hours of the evolution of SN\,2023ixf and discuss an appropriate model in Section~\ref{subsect:early}. Then we discuss stellar evolution scenarios and other phenomena that can explain the path of a massive star to an explosion that looks like SN\,2023ixf in Section~\ref{sect:evol}. We summarize our findings in Section~\ref{sect:conclusions}.

\section[Progenitor model and modeling of light curves]{Progenitor model and modeling of light curves} 
\label{sect:method}

\begin{table*}[!h]
\caption{SN LC models presented in our study.}\begin{center}
\resizebox{\textwidth}{!}{
\begin{tabular}{c|c|c|c|c|c|c|c|c|c|c|c|c|c|c|c|c|c|c}
\hline
\begin{tabular}[c]{@{}c@{}}Model\\name\end{tabular} & bare & W12& W13 & W14 & W15 & W16 &W17 &W5  &W1  &W2  &W3  &W4  &W6  & W9  & W8  &W7  &  W10 & W11 \\
\hline
\begin{tabular}[c]{@{}c@{}} $R_\mathrm{CSM}$ \\ $\left[ 10^{14} \,\mathrm{cm} \right]$ \end{tabular}& --- &\multicolumn{6}{c|}{1.5}&  \multicolumn{6}{c|}{6}  & \multicolumn{5}{c}{10}\\
\hline
\begin{tabular}[c]{@{}c@{}} $M_\mathrm{CSM}$ \\ \char`[  \Msun \char`] \end{tabular}& --- & 0.07& 0.17& 0.3& 0.55& 0.83& 1.56&0.07& 0.17& 0.3& 0.55& 0.85& 1.56&0.07& 0.17& 0.3& 0.55& 0.83\\
\hline
\begin{tabular}[c]{@{}c@{}} $\rho_\mathrm{in}$ \\ \char`[  $10^{-12}$\, g\,cm$^{-3}$ \char`] \end{tabular}& --- & 52.6 & 95.5 & 184.3 & 381.0 & 609.8 & 1176.4 & 5.7 & 14.2 & 27.3 & 52.6 & 95.5 & 184.3 & 3.0 & 6.9 & 14.2 & 27.3 & 52.6\\
\hline
\begin{tabular}[c]{@{}c@{}} correlation \\ coefficient \end{tabular}& --- & -0.45 & -0.44 & 0.09 & -0.09 & -0.32 & -0.22 & -0.27 & 0.40 & 0.68 & 0.69 & 0.58 & 0.40 & 0.56 & 0.79 & 0.82 & 0.92 & 0.91 \\\hline
\end{tabular} }
\end{center}\label{table:models}
\tablefoot{
With ``bare'' we denote the modified progenitor model without any appended CSM, i.e. a bare model (for details, see the text). ``W''-models are those which combine the bare progenitor with an appended CSM (``W'' stands for ``wind''). $R_\mathrm{CSM}$ is the outer radius of the wind-like CSM region, $M_\mathrm{CSM}$ the CSM mass, and $\rho_\mathrm{in}$ the density at the inner radius of the CSM region. The Spearman correlation coefficients for the $g$-band LCs, which we discuss in Section~\ref{sect:res}, are listed in the bottom row.\\
The numbering of the models corresponds to the sequence in which they were computed. }
\end{table*}

We utilised a non-rotating stellar evolution model s10 \citep{2016ApJ...821...38S} which has an initial mass of 10~\Msun{} and solar metallicity \citep[$Z_\odot=0.014$,][]{2005ASPC..336...25A}. This model is one of five baseline models in the model grid calculated by \citet{2023PASJ...75..634M}. The model provides the best fit for SN\,2023ixf according to \citet{2024PASJ...76.1050M}, since a relatively short plateau requires low-mass ejecta: the final pre-collapse mass of s10 is 9.7~\Msun{} ($M_\mathrm{ej}=8.3$~\Msun{}) and the radius is 490~\Rsun{}. In our study we modified the model in the following way: 
\begin{enumerate} 
\item We cut the outer layers of the stellar envelope up to the mass coordinate of 7~\Msun{} \citep[see also][]{2025ApJ...978...36F}. 
\item We stretched the star to a radius of 570~\Rsun{} while scaling the density profile and keeping the total mass of the system.
\item We appended an artificially constructed layer to force the density to drop to $10^{\,-11}$~g\,cm$^{\,-3}$, which leads to a surface radius, i.e. radius of the modified progenitor, of 700\,\Rsun{}.
\item We placed radioactive nickel $^{56}$Ni within 0.1~\Msun{} above 1.45~\Msun{}, which represents the inner boundary of the computational domain and defines the neutron star left behind in the core collapse. The mass of $^{56}$Ni is 0.05~\Msun{}. In our study we did not adjust the mass of $^{56}$Ni to match the tail luminosity, since there is no big effect on the duration of the plateau and its luminosity, if the $^{56}$Ni mass varies within 10--20\,\% around 0.05~\Msun{} \citep{2009ApJ...703.2205K,2016ApJ...821...38S,2019MNRAS.483.1211K}. We also did not consider any mixing of $^{56}$Ni, however, it may influence some of the predicted characteristics of LCs, such as colours.
\end{enumerate} 
At the end of the modifications our model, i.e., the central neutron star ($\sim$1.45\,M$_\odot$) plus the ejecta, had a total mass of 7.4~\Msun{}. We chose an explosion energy of 0.7~foe, which is injected as thermal energy into the innermost 0.6~\Msun{} of the ejecta. Our motivation for this particular amount of energy is explained in Section~\ref{sect:intro}. We call this modified model ``bare'' in our discussion below.
In Appendix~\ref{appendix:append1}, we provide a more detailed description of all modifications including corresponding plots.

In all other investigated models we appended a wind-like CSM with $\rho\propto r^{-2}$ to the constructed progenitor. Note that we did not consider density gradients other than power-laws with exponent $-2$, and we did not conduct a full parameter study but limited our parameters to the values published in literature. We chose three values for the radius of the CSM: (1) $1.5\times10^{14}$~cm, 2\,150~\Rsun{}; (2) $6\times10^{14}$~cm, 8\,630~\Rsun{}; and (3) $1\times10^{15}$~cm, 14\,380~\Rsun{}. While the density where the CSM is attached to the stellar profile varies, the total mass of the CSM ranges between 0.07~\Msun{} and 1.56~\Msun{}. The velocity of the CSM was set to zero, since we did not see a visible effect on the resulting LCs in test calculations with assumed velocities of 10~km s$^{-1}$ and 100~km s$^{-1}$. The parameters of our models are listed in Table~\ref{table:models}. The models there with capital letter ``W'' denote our cases with wind-CSM, and model ``bare'' means the modified s10 progenitor without any CSM. In Appendix~\ref{appendix:append2}, we show a few of our CSM models for illustrative purpose.

\subsection[Light curve modelling]{Light curve modelling} 
\label{subsect:method2}

We followed the SN explosion and modeled the LCs with the one-dimensional (1D) multi-band radiation hydrodynamics code \STELLA\, \citep[][]{1998ApJ...496..454B,2000ApJ...532.1132B,2006AandA...453..229B}. A description of \STELLA, including, particularly, details about opacities, and comparison to other radiative transfer codes is presented in \citet{2022A&A...668A.163B}.


\section[Results and Discussion]{Results and Discussion} 
\label{sect:res}

In Figures~\ref{figure:lbol} and \ref{figure:pseudo}, we present the bolometric and pseudo-bolometric LCs for our models, respectively. Our bolometric LC is truly bolometric. The LC is therefore an integration of the flux over the full simulated wavelength (frequency) range, from 1~\AA{} to 50\,000~\AA{}. The pseudo-bolometric LC is an integration of the flux over the standard Bessel broad bands $U$, $B$, $V$, $R$, and $I$. Bolometric LCs of SN\,2023ixf taken from the literature do not necessarily cover the entire electromagnetic range and usually represent an approximate estimate based on a certain procedure, e.g., integration of the flux over the available spectral broad bands and extrapolation to the longer and shorter wavelengths with a black-body distribution \citep{2024A&A...683A.154M}. \citet{2024Natur.627..759Z} presented a bolometric LC which includes the observed UV photometry, i.e. it can be considered close to truly bolometric. 
We note that \citet{2024A&A...683A.154M} constructed their LC while implementing bolometric corrections, which leads to an overestimated sharp peak in the bolometric LC not present in any other studies.
``Observed'' pseudo-bolometric LCs are calculated differently in different studies, e.g., \citet{2024ApJ...975..132S} integrated the flux over $UVOIR$ bands. This means that comparing our synthetic LCs to the ``observed'' ones in that study might be misleading. Nevertheless, we still do the comparison and draw some conclusions.

In our figures, we collected data for SN\,2023ixf from different studies: \citet{2023ApJ...955L...8H,2023ApJ...954L..12T,2024ApJ...975..132S,2024A&A...681L..18B,2024A&A...683A.154M,2024Natur.627..754L,2024arXiv240807874H}. We choose MJD=60\,082.788 as a reference point \citep{2024Natur.627..754L}, which indicates time ``0'' in all figures in our paper. The so-called ``explosion epoch'' or ``time of explosion'' for SN\,2023ixf is defined as the average between the first detection and the last non-detection and corresponds to MJD\,=60\,082.75 \citep{2023ApJ...953L..16H}. Other ways of estimating the ``explosion epoch'' include fitting of the early LC with a power-law and choosing the ``explosion time'' as the time when the flux equals zero \citep{2023SciBu..68.2548Z}. Nevertheless, these methods, being dependent on the limiting magnitude or epoch of the last non-detection, do not necessarily reflect the physical meaning of the time of the actual explosion. 
To be physically precise, the ``explosion epoch'' corresponds to the onset of an explosion, i.e. the moment when the SN shock is successfully launched in the vicinity of a stellar core, starting its acceleration onwards. The shock propagates to the edge of the star during $t_\mathrm{sbo}=R_\mathrm{prog}/v_\mathrm{sh}$, where $v_\mathrm{sh}\sim 10\,000$~km\,s$^{\,-1}$ is the shock velocity, which equals 0.6~days for a progenitor similar to our ``bare'' model with $R_\mathrm{prog}=700$\,\Rsun{}. It means that the first photons can be detected 0.6~days after the explosion, i.e. the actual time of the explosion and observationally derived ``time of explosion'' might be different.
By default, time ``0'' in \STELLA\, simulations corresponds to the moment when the thermal energy is injected in the innermost layers of the ejecta. This means that it is close to the actual beginning of the explosion. 
Our CSM$-$interaction models consist of an artificially constructed CSM and therefore the models require some time to adjust to an equilibrium structure according to the underlying set of equations. This means that the first day or so in the LC evolution might be unphysical. Therefore, there is some degree of freedom to shift different models to a different time, possibly within $2-11$~days, to match the observations, in which one to two days account for the nonphysical part of a synthetic LC.

\subsection[LCs of the bare progenitor]{LCs of the bare progenitor} 
\label{subsect:bare}

\begin{figure*} 
\centering 
\hspace{-5mm}\includegraphics[width=\textwidth]{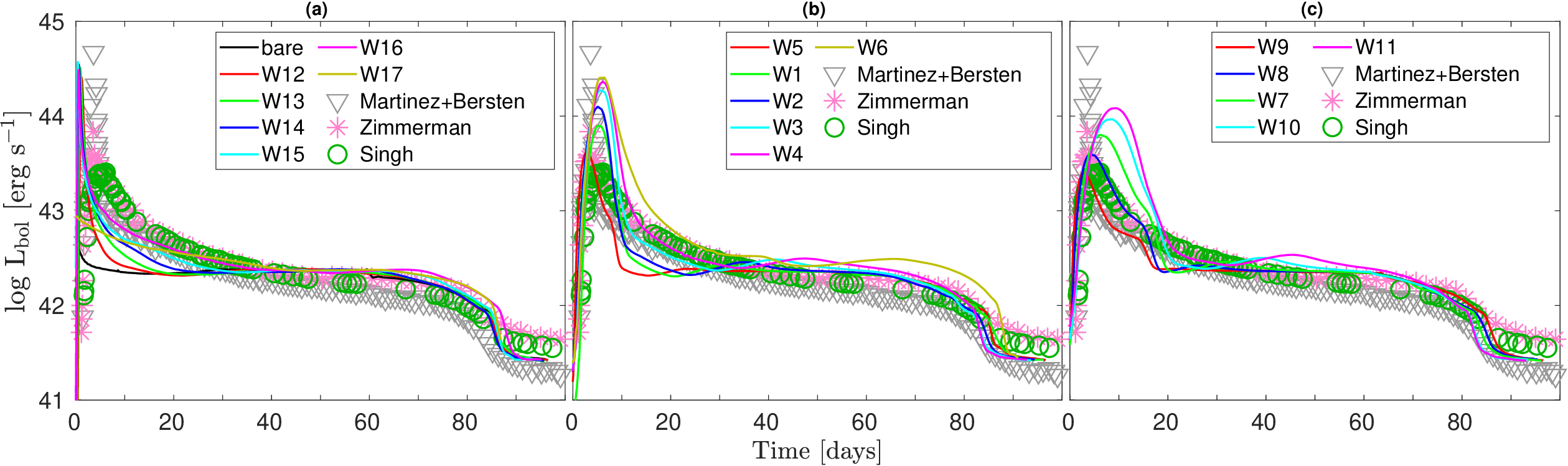} 
\caption{Bolometric LCs for a grid of models with CSM radii: $1.5\times10^{14}$~cm (a; left), $6\times10^{14}$~cm (b; middle) and $10^{\,15}$~cm (c; right). The observational data are taken from \citet{2024A&A...681L..18B}, \citet{2024A&A...683A.154M}, \citet{2024ApJ...975..132S}, and \citet{2024Natur.627..759Z}.} 
\label{figure:lbol} 
\end{figure*}

\begin{figure*} 
\centering 
\hspace{-5mm}\includegraphics[width=\textwidth]{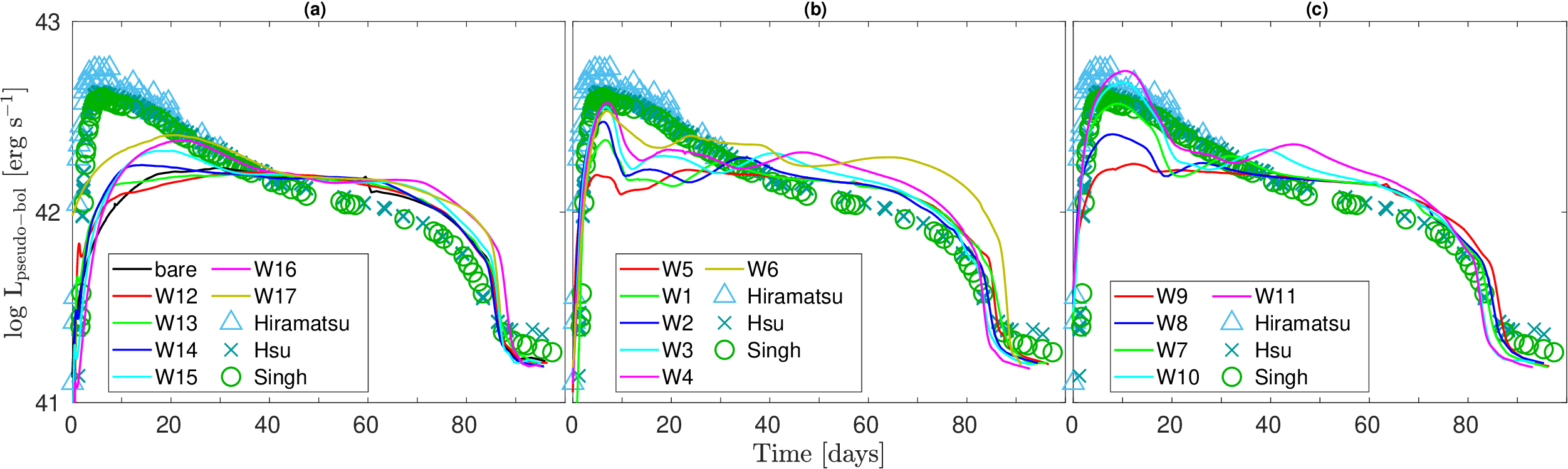} 
\caption{Pseudo-bolometric LCs for a grid of models with CSM radii: $1.5\times10^{14}$~cm (a; left), $6\times10^{14}$~cm (b; middle) and $10^{\,15}$~cm (c; right). The observational data are taken from \citet{2023ApJ...955L...8H}, \citet{2024arXiv240807874H}, and \citet{2024ApJ...975..132S}.} 
\label{figure:pseudo} 
\end{figure*}

Figures~\ref{figure:lbol}a and \ref{figure:pseudo}a show the LC for the bare progenitor, and we begin with a discussion of the properties of this LC.

Our goal was to build a progenitor and simulate its explosion with an energy that is consistent with the explosion energies derived from observations of SN remnants and with the neutrino-engine-driven and first-principle multi-dimensional CC simulations that we mentioned in Section~\ref{sect:intro}. The resulting LC of our bare model, which does not have a CSM, with an energy of 0.7~foe matches the plateau phase of SN\,2023ixf LC. 
Regarding the ejecta mass and the explosion energy, our result is in good agreement with the conclusions of \citet{2024arXiv240807874H}, who explained SN\,2023ixf as an explosion of a massive star above 15~\Msun{} that underwent an intense mass-loss phase and massive fallback, retaining a low mass of hydrogen-rich ejecta, and that had an energy of $0.5-0.7$~foe. The progenitor radius in their self-consistent simulations, which followed the progenitor variability, is larger than the radius we set artificially in our input model, namely $\geq 950$\,\Rsun{} instead of $700$\,\Rsun{} for our model. Moreover our progenitor model is not a self-consistently calculated stellar evolution model, therefore, we cannot also deduce the initial mass of the progenitor. These are weak points in our study. The lower ejecta mass in \citet{2024arXiv240807874H} simulations was achieved also because of the high fallback mass, up to 3.6 ~\Msun{}, happening after the successful explosion. In our simulations, there is no matter falling back to the gravitational central object. The reason is most likely a different density structure of the inner regions lying above the collapsing core.

With our experiment of reducing the energy of the explosion in combination with lowering the ejecta mass and extending the envelope, we show that the plateau luminosity and its duration can be similar to the LC obtained with a higher energy of 2--3~foe and a higher ejecta mass \citep{2023ApJ...954L..12T,2024PASJ...76.1050M,2024ApJ...975..132S}. 
Hence, instead of considering such high values, the LC of SN\,2023ixf can also be explained by an explosion with a more common energy of less than 1~foe, which is more consistent with well-accepted values predicted by self-consistent neutrino-driven CCSN simulations and derived from observations.

The degeneracy of bolometric LCs of Type\,IIP was explored by \citet{2019ApJ...879....3G}. The same behavior of the bolometric LC can be reproduced by progenitors with significantly different structures and masses. E.g., the ejecta mass and explosion energy may differ by a factor of two and still reproduce the same shape of a bolometric LC, which is consistent with the scaling relations by \citet{1980ApJ...237..541A} and \citet{1993ApJ...414..712P}:
\begin{equation}
\begin{array}{l}
\log L \sim - 1/2\, \log M - 0.8\, \log E + 2/3\, \log R \\
\log t_\mathrm{p} \sim 1/2\, \log M - 1/6\, \log E + 1/6\, \log R\, .
\label{popov}
\end{array}
\end{equation}

The higher the ejecta mass the longer the plateau, the higher the energy the shorter the plateau, the larger the progenitor radius the longer the plateau; and vice versa: the lower the ejecta mass the shorter the plateau, the lower the energy the longer the plateau, and the smaller the radius, the shorter the plateau. In particular, the bolometric LC of SN\,2023ixf can be a result of a high energy in combination with a high ejecta mass, or the other way around: a lower energy and a lower ejecta mass can yield the same result according to Equation~\ref{popov}. Moreover, an estimate of the progenitor and CSM parameters might be misled if one uses either only bolometric or only pseudo-bolometric LCs. Strong constraints on the radius of the progenitor can break the degeneracy, which can be done, e.g., by considering the colour-temperature evolution before the LC settles to the recombination phase \citep{2010ApJ...725..904N,Shussman2016,2020MNRAS.494.3927K}. Spectral modelling either for the photospheric phase or the nebular epoch definitely provides an additional constraint on the progenitor parameters \citep{2012A&A...546A..28J,2013MNRAS.433.1745D,2023A&A...677A.105D,2024A&A...687L..20F}. 

In conclusion, we managed to reproduce the LC of SN\,2023ixf, namely, the duration of the plateau and its luminosity after the ``interaction'' phase with an explosion powered by 0.7~foe.

\subsection[LCs of the progenitor with CSM]{LCs of the progenitor with CSM} 
\label{subsect:lcs}

In Figures~\ref{figure:lbol} and \ref{figure:pseudo}, we show the bolometric and pseudo-bolometric LCs for the progenitor enshrouded by CSM extending to different radii: $1.5\times10^{14}$~cm (models W12--W17), $6\times10^{\,14}$~cm (models W1--W6) and $10^{15}$~cm (models W7--W11). The mass of this CSM ranges between 0.07~\Msun{} and 1.56~\Msun{} (see details in Table~\ref{table:models}).

The model with a compact CSM can reproduce the bolometric peak but fails to reproduce the slope of the rising part of the SN\,2023ixf LC. The CSM cases with a radius of $6\times10^{14}$~cm and masses of 0.07~\Msun{} or 0.17~\Msun{} resemble the peak shape, but drop at day~10 and last not long enough, while the CSM cases with a radius of $10^{\,15}$~cm and masses of 0.07~\Msun{} and 0.17~\Msun{} match the peak and its width. The higher-mass CSM overestimates the luminosity of the peak and its width, because the diffusion time is longer for the higher ejecta mass. The preferable models are: W1, W2 ($6\times10^{14}$~cm, 0.17~\Msun{} and 0.3~\Msun{}), and W8 and W7 ($10^{15}$~cm, 0.17~\Msun{} and 0.3~\Msun{}), if we base our analysis on the bolometric LC. \citet{2024A&A...683A.154M} inferred a CSM mass of 0.02~\Msun{} while fitting the shape of their bolometric LC, although their LC has a very sharp, probably unrealistic, peak and is an outlier among others.

The rising part of the LCs corresponds to the phase when the shock wave propagates through the CSM. As the shock is radiation-dominated, it accelerates particles via radiation pressure ahead, developing the so-called ``thermal tongue'' \citep{1967pswh.book.....Z,1971Ap&SS..10...28G,1976ApJ...207..872C}. 
Radiative acceleration pushes matter ahead of the shock, which leads to the emergence of the flash-ionisation features.

While looking at pseudo-bolometric LCs (Figure~\ref{figure:pseudo}), it is obvious that none of the models with the compact CSM matches the observations because of a flux deficit before day~20, which accounts for interaction. The models with $R_\mathrm{CSM}=6\times10^{14}$~cm can be considered matching the peak pseudo-bolometric luminosity of SN\,2023ixf if the CSM mass is 0.55~\Msun{}, 0.85~\Msun{} or 1.56~\Msun{}. However, the slope of the rising part is shallower and the ``interaction'' bump is narrower than that of SN\,2023ixf. The models with $R_\mathrm{CSM}=10^{15}$~cm and CSM mass of 0.3~\Msun{}, 0.55~\Msun{}, and 0.83~\Msun{} match the pseudo-bolometric LC of SN\,2023ixf reasonably well.

There is some degree of discrepancy between the estimate of the CSM mass via bolometric LCs and pseudo-bolometric LCs. The lower mass of the CSM required to match the bolometric luminosity may be explained by the fact that our synthetic bolometric LC is an integration of the flux over the entire electromagnetic spectrum, whereas the ``observed'' bolometric luminosity is a sum of the fluxes in a certain set of broad bands with additional components such as bolometric corrections and other adjustments. Therefore, the estimate based on the pseudo-bolometric LC is more appropriate, although the comparison in broad bands can be more reliable, and we discuss this below.

\subsection[Broad-band LCs]{Broad-band LCs} \label{subsect:bands}

In Figures~\ref{figure:uvw2_all}, \ref{figure:u_all}, \ref{figure:V_all}, \ref{figure:g_all}, \ref{figure:r_all}, and \ref{figure:z_all} (see Appendix~\ref{appendix:bands}), we show broad-band LCs for our models. As concluded above, the models with the compact CSM are ruled out because of the flux deficit at earlier time (before day~20), which is also seen in broad-bands. Therefore, we show models W12--W17 in Figures~\ref{figure:g_all} and \ref{figure:r_all} only for illustrative purposes. Otherwise, we concentrate our analysis on our models W1--W11.

The plateau part of the LCs matches the observed LC of SN\,2023ixf reasonably well, at least not worse than, e.g., in \citet{2024A&A...681L..18B} or \citet{2024PASJ...76.1050M}. Although the flux in the $u$-band an $z$-band is overestimated in all models on the plateau compared to the observed one in SN\,2023ixf, agreement in broad bands is acceptable. This means that our progenitor model, which determines the behaviour of the plateau, is suitable to mimic the structure of the star before the CC.

For the interaction part, none of the CSM structures can reproduce the behaviour of SN\,2023ixf in all broad bands simultaneously. However, there are certain preferences for models: W10 and W11 match the observed LCs in a larger number of broad bands, which means that more extended, $10^{15}$~cm, and more massive, $0.55-0.85$~\Msun{}, CSM cases suit SN\,2023ixf better. Models W4, W7, W8, and W9 are deemed to be appropriate, too. 
We note that studies by, e.g., \citet{2023ApJ...955L...8H} based their analysis on pseudo-bolometric and bolometric LCs. Using such data, however, can be ambiguous and therefore not finally discriminating for the best-fit case because of the inherent difficulties in comparing wave-band integrated data from observational analysis with results from theoretical LC modeling.

To justify our choice of the best-fit models, we conducted a statistical analysis. We chose the Spearman correlation method to differentiate between models. To do so, we combined all available observational data in each band into a single array, and calculated Spearman correlation coefficients for each model in all bands. We processed all models within a time window between day~0 and day~30, without considering correlations for the ``plateau'' phase, because the progenitor model was built specifically to well reproduce the plateau length and luminosity. Models W2, W3, W7, W8, W10, and W11 have the largest correlation coefficients, between 0.7 and 1. This result agrees with our assessment we made by simple eye inspection. 

None of the models has sufficient flux in the $z$-band to reproduce the $z$-magnitude of SN\,2023ixf during the first 30~days.
The underestimated flux in red bands in combination with overestimated flux in blue bands can indicate that the colour temperature in our models is somewhat too high to explain the colours of SN\,2023ixf. It might be that a slightly higher metallicity in our input progenitor could suppress the blue flux and redistribute it to the redder part of the spectrum. Another possibility might be connected to the degree of mixing of radioactive nickel $^{56}$Ni, according to which the SN ejecta at larger radii have a higher fraction of nickel in comparison to centrally-concentrated nickel. Nickel together with other iron-group elements has large line opacities, which in turn, effectively redistribute blue flux into redder wavelengths \citep{2000ApJ...530..757P,2006ApJ...649..939K,2020MNRAS.499.4312K}.


\subsection[The first five hours of SN\,2023ixf]{The first five hours of SN\,2023ixf} 
\label{subsect:early}

\begin{figure} 
\centering 
\hspace{0mm}\includegraphics[width=0.48\textwidth]{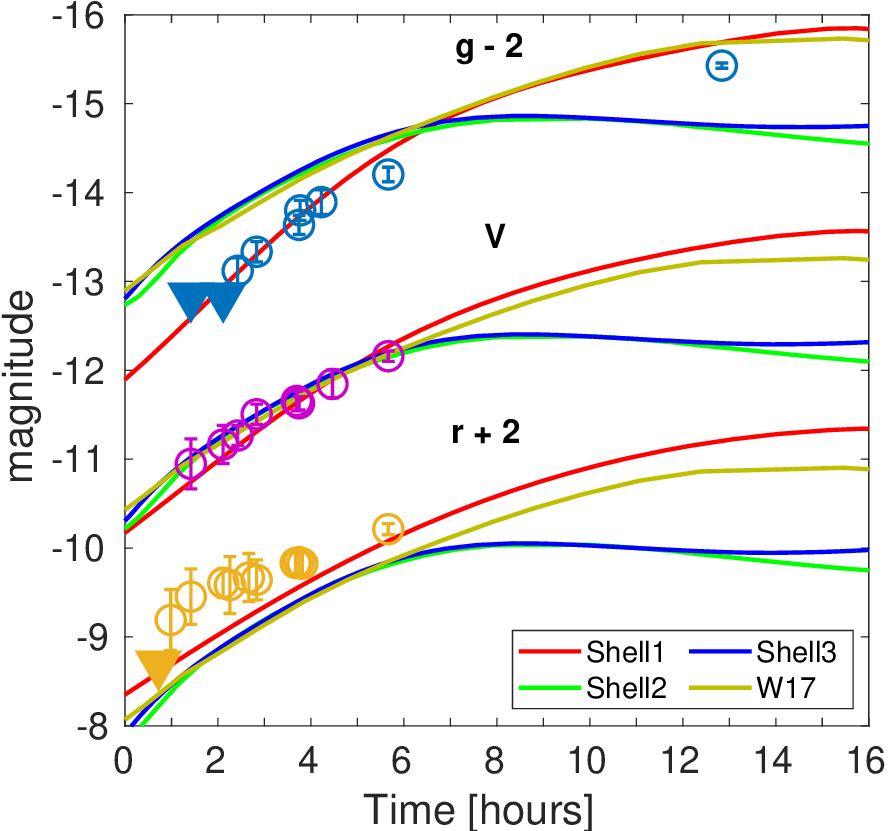} 
\caption{$g$, $V$, and $r$ LCs for a CSM-model W17 with a compact CSM ($1.5\times10^{\,14}$~cm, 1.56~\Msun{}), and shell-models: ``Shell1'', ``Shell2'', and ``Shell3'' with an inner radius of 700~\Rsun{}, outer radii of 1000~\Rsun{}, 900~\Rsun{}, and 800~\Rsun{}, masses of 0.9~\Msun{}, 1~\Msun{}, and 0.5~\Msun{}, and injected energies of 0.01~foe, 0.02~foe, and 0.0185~foe, respectively. The observational data are taken from \citet{2024Natur.627..754L} and are shown by circles. The triangles indicate non-detection and are also from \citet{2024Natur.627..754L}.} 
\label{figure:bump} 
\end{figure}

\citet{2024Natur.627..754L} achieved to capture emission from SN\,2023ixf already during the first hours after confirmed non-detection (Figure~\ref{figure:bump}) and 22.5 hours, i.e. almost one day, before the discovery observation of the SN  
by \citet{2023TNSTR..39....1I}. These observations are unique and provide extremely valuable information of the very compact CSM surrounding the surface of the progenitor or of an outermost, puffy atmosphere of the progenitor. Therefore they indicate an unusual dynamical activity happening a few hours before or after the CC. Therefore, we pay special attention to the first five hours of SN\,2023ixf LC and analyse this phase in more detail.

Our best-fit models, i.e., those with CSM of $6\times10^{14}$~cm and $10^{15}$~cm, have a slope that is shallower than SN\,2023ixf and slightly overestimates the flux in this earliest epoch, which is not seen in Figures~\ref{figure:lbol}, \ref{figure:pseudo}, and Figures~\ref{figure:uvw2_all}--\ref{figure:z_all} because the plots cover 100~days of the LC evolution. Models W12--W17 with the more compact CSM of $1.5\times10^{14}$~cm have a sharper rise to the SN LC peak but suffer from the flux deficit around the peak. Our model W17 with a CSM mass of 1.56~\Msun{} extending to $1.5\times10^{14}$~cm reproduces the earlier LC of SN\,2023ixf.  However, 1.56~\Msun{} of CSM in this case have a pronounced impact on the plateau duration, making it 10~days longer\footnote{We note that the LC of W17 is shifted by 11~days in Figures~\ref{figure:lbol}, \ref{figure:pseudo}, and Figures~\ref{figure:uvw2_all}--\ref{figure:z_all} to match the plateau phase of SN\,2023ixf.}.
The immediate explanation of SN\,2023ixf then can be that the progenitor was not surrounded with a single type of CSM but had an asymmetric density profile, e.g., more extended along one axis and less extended along another axis. 
An asymmetric compact CSM around the progenitor of SN\,2023ixf was in fact confirmed by spectropolarimetry observations \citep{2023ApJ...955L..37V,2024ApJ...975..132S}. It might be that a combination of different CSM structures contributes to the overall LC gradually, in other words, photons from the shock propagating through the more compact CSM break out earlier, then photons from the shock propagating along the more extended CSM dominate the LC a few hours later.
We also tried more compact CSM cases with radii of $10^{\,14}$~cm and smaller and lower masses of $0.001-0.01$~\Msun{}. However, such structures of the CSM lead to a steeper slope and thus cannot be considered as successful in reproducing the early phase of the LC of SN\,2023ixf. 

In any case, the early LC rise, which appeared like a flux excess in the analysis of \citet{2024Natur.627..754L} cannot be satisfactorily reproduced by any of our tested SN models with CSM. Instead, in 1D models it can only be explained by an additional thin layer or shell that lies right on top of the progenitor and has a substantially higher density than the ambient CSM (the density at the inner edge of the CSM is between $\sim$$3\times 10^{-12}$\,g\,cm$^{-3}$ and $10^{-9}$\,g\,cm$^{-3}$, whereas the density at the inner edge of the shells we will consider below is taken to be $10^{-8}$\,g\,cm$^{-3}$). We note that shocking such a layer with an energy equivalent to the explosion energy will lead to a very sharp rise to a high-luminosity feature. In order to reduce this luminosity spike, the thin shell or a compact CSM should be unreasonably massive, far above 1~\Msun{}, like our W17 model with the CSM mass of 1.56~\Msun{}. The higher mass would make the rise more shallow, however, meaning that the LC is less suitable for matching the early rise. This is in accordance with the inverse dependence of the shock-breakout luminosity on the mass and its linear dependence on energy. Thus, we conclude that the energy powering the 5-hour feature has to be lower than 0.7~foe.

\begin{table}[!h] 
\caption{Shell-model parameters.} \begin{center}
\begin{tabular}{p{0.1\textwidth}<{\centering} p{0.1\textwidth}<{\centering} p{0.1\textwidth}<{\centering} p{0.1\textwidth}<{\centering} } \hline
models & M$_\mathrm{sh}$ [\Msun{}] & R$_\mathrm{sh}$ [\Rsun{}] & E$_\mathrm{inj}$ [foe] \\     \hline
``Shell1'' & 0.9 & 700--1000 & 0.01 \\
``Shell2'' & 1.0 & 700--900  & 0.02 \\
``Shell3'' & 0.5 & 700--800  & 0.0185 \\ \hline
\end{tabular}  
\end{center}\label{table:shells-param}
\tablefoot{
$M_\mathrm{sh}$ is the mass of the shell in solar masses, $R_\mathrm{sh}$ is the radial range of the shell, where the first number means the inner boundary and the second number represents the outer boundary of the shell in solar radii, and $E_\mathrm{inj}$ is the energy injected at the inner boundary of the shell in units of foe.}
\end{table}

Therefore, we tested another hypothetical scenario, assuming that the additional shell around the progenitor radiates energy in an expansion that is driven by much less energy. To this end we calculated a set of independent models as an alternative attempt to reproduce the initial LC rise as required by the observations in the time window of five hours before the first official discovery of SN~2013ixf.
Specifically, we constructed a variety of shell models with masses between 0.5~\Msun{} and 1~\Msun{}, inner radius of 700~\Rsun{}, and outer radii of 800~\Rsun{}, 900~\Rsun{}, and 1000~\Rsun{}. To build these shell models, we used the outer mass zones of the progenitor structure (our base model ``bare'') including their chemical composition and scaled the density and radius to the desired values, while keeping the inner density at $10^{\,-8}$\,g\,cm$^{\,-3}$.
We mapped the shell models into \verb|STELLA| and followed their evolution after injecting energies between 0.0005~foe and 0.02~foe, i.e., $5\times10^{47}$~erg and $2\times10^{49}$~erg, into the innermost computational grid zones of the shells.
In Table~\ref{table:shells-param} we list the best-fit cases, i.e., only those combinations which, from our point of view, yield the best match of the first 5-hour phase of SN\,2023ixf.

In Figure~\ref{figure:bump}, we present the resulting LCs of the expanding shells. We note that these models do not include the combined system of ``progenitor+shell+CSM'', but consist exclusively of a shell that is gravitationally unbound from the progenitor. Additionally, we show the CSM-model W17, which has a LC appropriate for SN\,2023ixf (see above). The radiation in the shell models is powered by the injected energy, which triggers a shock propagating across the shell and causes radiation streaming away from the shock front.
The synthetic LCs in $g$, $V$, and $r$ broad bands are shown in comparison to those of SN\,2023ixf. 
All shell-models and W17 reproduce the $V$-band LC of SN\,2023ixf, while suffering a flux excess in the $g$ band and a flux deficit in the $r$ band. Nevertheless, our W17 wind-model includes 1.56~\Msun{} of CSM within $1.5\times10^{14}$\,cm, which we consider being too massive to be realistic.
It might be that the large amount of dust reported by \citet{2023ApJ...955L..15N} suppresses flux in the bluer bands such as the $g$-band  \citep[see also][]{2023ApJ...952L..23K, 2023ApJ...953L..14P, 2024MNRAS.534..271Q, 2024ApJ...968...27V, 2024SCPMA..6719514X}.
In addition to Figure~\ref{figure:bump}, we show $V$-band LCs for our shell-models together with the best-fit wind-models 
in Appendix~\ref{appendix:append3}.

It would overstate the case if we claimed that our models provide a good fit to the data, 
but the shell models at least follow the LC slopes in the early phase of SN\,2023ixf.  
Therefore, we conclude that a small part of the outermost envelope of the progenitor could have been swept out by a burst of energy of about 0.01--0.02~foe a few hours before the main shock from the explosion reached the edge of the progenitor. The mass ejection of the shell could also be asymmetric. The proposed shell ejection is in agreement with other studies on SN\,2023ixf claiming that the progenitor's mass-loss rate increased rapidly while the star was approaching its CC \citep{2024ApJ...973L..47B}.

In our hypothetical scenario, we assume that an additional energy deposition happens in the outermost layers of the progenitor almost simultaneously with the main explosion, i.e., shortly before the breakout of the SN shock from the stellar surface. If the corresponding injection of a low amount of energy occurs at a smaller radius somewhere in the interior of the pre-collapse star, the weak shock formed by that will propagate through the rest of the envelope only over days \citep[see][]{2024arXiv241007055T}. This means that a close temporal correlation of shell ejection and SN outburst would require quite some fine-tuning in order to avoid that the shell is expelled much too early or that there may not be enough time for the weak shock to reach the edge of the stellar envelope before the main shock from the explosion catches up with it.

In the next section we discuss some possible origins of the assumed pre-SN outburst and its corresponding injected energy, and the connection to the required CSM around the progenitor of SN\,2023ixf.


\section[Origin of the CSM around SN\,2023ixf progenitor and origin of the pre-SN outburst]{Origin of the CSM around SN\,2023ixf progenitor and origin of the pre-SN outburst}
\label{sect:evol}

As discussed above, there are three important ingredients required to reconstruct the LC appearance of SN\,2023ixf and, in turn, to explain the physics and nature of this transient. These are: (1) the progenitor structure, namely, its final mass just before the CC, which is about 7.4~\Msun{} in our model --\,and, corresponding to it, the ejecta mass of about 6~\Msun{}\,-- as well as its radius of about 700~\Rsun{}; (2) the CSM structure: a mass of about 0.55~\Msun{} with a radius of $10^{\,15}$~cm; and (3) the shell of stellar material ejected with energy of about 0.01~foe to explain the steep initial rise of the LC during the first five hours. In the following, we attempt to sketch possible scenarios to fulfil these requirements.

\subsection[Binary nature]{Binary nature}
\label{subsect:binary}

To match the plateau phase of SN\,2023ixf, we constructed a progenitor with a final mass of 7.4~\Msun{}, which is lower than the final mass of stellar models with an initial mass of $10-20$~\Msun{} at solar metallicity \citep{2004A&A...425..649H}. This means that the progenitor had to lose more mass than the steady wind mass-loss predicts. One of the most probable channels is to consider a star within a binary system so that the outermost layers are transferred via Roche-Lobe overflow (RLOF) to a companion \citep{2021A&A...656A..58L,2022A&A...662A..56K,2024A&A...685A..58E}. The mass transfer rate can be so high that the companion is not able to accrete this amount of mass, and a circumbinary disc is formed, in turn, extending to radii a few times the orbital separation.  
It is well known that extra mass loss due to binary interaction plays a significant role for stripped-envelope SNe of Type\,Ibc \citep{2012A&A...544L..11Y,2022ApJ...929L..15F}.

\citet{2024arXiv240807874H} discussed that there should be two processes or episodes that led to intensive mass loss of the progenitor system of SN\,2023ixf: the first episode is responsible for stripping the progenitor to reduce its mass during the earlier phases of evolution, and the second episode via which the CSM was formed. The progenitor is supposed to lose a high fraction of its hydrogen-rich envelope during earlier phases of evolution. For example, if a non-rotating star is 15~\Msun{} initially, it loses 2.4~\Msun{} via wind mass loss \citep{2016ApJ...821...38S}. However, the progenitor of SN\,2023ixf had to have 7.4~\Msun{} before its CC, and assuming it was 15~\Msun{} at the beginning, it had to lose 7.6~\Msun{}, which means that an extra 5~\Msun{} should be removed during its evolution, including the red supergiant (RSG) phase. Similarly, a 10~\Msun{} star loses 0.3~\Msun{} via its wind and is required to lose another 2.3~\Msun{} to be a suitable progenitor for SN\,2023ixf. As shown in the literature, increasing the mass-loss rate by a fixed factor might not necessarily help \citep{1994A&AS..103...97M}, since the population of RSGs and hydrogen-depleted stars will not be consistent with observations. 

To conclude, it is likely that the progenitor of SN\,2023ixf was part of a binary and had lost its mass both via wind mass-loss and binary interaction. The CSM may represent the matter that escaped from the primary during core carbon, neon, and oxygen burning in a dynamically unstable regime, or in a phase of super-thermal accretion by the secondary (i.e., the mass transfer from the primary happened at a high rate on a time scale shorter than the Kelvin-Helmholtz, or thermal, time scale). This matter remained trapped in the system and could form the CSM \citep{1977A&A....54..539K,2024ApJ...966L...7L}. We discuss more details of some possible CSM formation scenarios in the next section.

\subsection[Pre-SN outburst and origin of CSM]{Pre-SN outburst and origin of CSM}
\label{subsect:preouburst}

According to our analysis, the steep rise in the SN\,2023ixf LC during the first hours after confirmed non-detection (see Figure~\ref{figure:bump}) might point to a low-energy pre-SN outburst. Our approximate model considers a shell of matter ejected from the progenitor about half a day before the shock breaks out from the surface of the progenitor. This outburst could be correlated with the SN or of uncorrelated nature. 

Uncorrelated nature means that the CC and the mass ejection happen independently of each other and are not connected. In this case, the mass ejection is a result of dynamical activity going on in extended RSGs. On the one hand, these stars possess convective envelopes with convective plumes encompassing the entire hydrogen-rich envelope of the star \citep{2009A&A...506.1351C,2018A&A...610A..29K,Chiavassa2022,2022ApJ...929..156G,2023ApJ...954L...1S}. On the other hand, sound waves and gravity waves arising from convective burning activity in the progenitor's core might transmit an energy of $10^{\,48}$~erg into the envelope during the final years before the CC \citep{2012MNRAS.423L..92Q,2021ApJ...906....3W}. This epoch corresponds to the nuclear burning stages of core-neon and core-oxygen burning, which last a few years. The thermal timescale of the envelope is the Kelvin-Helmholtz time and is equal to a few hundreds of years, meaning that the envelope cannot respond immediately to nuclear and dynamical processes in the core evolving on a shorter timescale. The energy released via core and shell burning is the source of energy for exciting the internal gravity waves. Merging sound waves and damped gravity waves result in outgoing acoustic waves and ultimately in shock waves that could initiate an expansion of the envelope or/and an ejection of outermost layers \citep{2016MNRAS.458.1214Q,2017MNRAS.470.1642F,2024OJAp....7E..47F} and the formation of an extended CSM. We note that such a long-lasting, extensive mass-loss phase over several final years of the progenitor's evolution might provide the CSM needed to explain the main LC of SN\,2023ixf.
Regarding the LC feature that is observed during the first hours, mass ejection caused by stochastic processes such as convection or strong gravity-wave activity might also be responsible. However, we emphasise that there would have to be a particular shell ejection just half a day before the main shock breaks out.

A mass ejection as an extreme case of an intensive mass-loss during the latest evolutionary stages can be a result of an amplified regular wind mass-loss.
The known RSG wind mass-loss rates are about $10^{-6}$~\Msun{}\,yr$^{\,-1}$ \citep{2018MNRAS.475...55B}, which is not sufficient for the formation of the CSM needed for SN\,2023ixf. However, wind mass-loss can be increased by pulsations \citep{1983ApJ...272...99W,1992ApJ...399..672W,1997A&A...327..224H,2010ApJ...717L..62Y}, which massive stars are predisposed for, and the SN\,2023ixf progenitor, too \citep{2023ApJ...952L..23K,2023ApJ...957...64S,2024arXiv240807874H}.

Alternatively, the mass ejection might also be correlated with the stellar collapse taking place in the deep interior of the stellar core. In Section~\ref{subsect:preouburst}, we considered shell models that match the early SN\,2023ixf LC. These models include a shell with a mass of $0.5-1$~\Msun{} and a width of $100-300$~\Rsun{} ejected with the energy of about 0.01~foe.

Our numbers for mass ($\sim0.5$~\Msun{}) and energy (0.01~foe) of the pre-SN outburst are consistent with results deduced from observations \citep{2021ApJ...907...99S} and a theoretical approach \citep[e.g.,][]{2022ApJ...936..114M}. However, they represent upper limits for these parameters, since the eruption might be asymmetric. It is possible that another outburst with a lower energy and lower luminosity occurred before the detection of SN\,2023ixf and remained undetected, because several studies reported non-detection of outbursts at the location of SN\,2023ixf for many years before the SN was detected \citep{2023RNAAS...7..174F,2024MNRAS.527.5366N,2024ApJ...965...93R}. There are SNe for which precursor emission was detected years before the terminal explosion \citep[see e.g.,][]{2022ApJ...924...15J}. For instance, \cite{2022MNRAS.512.2777T} suggested that a CSM around SN\,2020fqv originates from an outburst that happened 200~days before the CC with energy of $5\times10^{\,46}$~erg, which is three orders of magnitude lower than our estimate. The matter stripped off during years prior to the CC indeed can be responsible for the formation of a CSM required for shaping the appearance of SN\,2023ixf, but not for the matter ejected within a fraction of day before the explosion.


The shock formed in the shell reaches the outer edge in 1, 0.6, 0.4~days for our models ``Shell1'', ``Shell2'', and ``Shell3'', correspondingly. This means that the shells did not have time to expand, keeping its original radius. By this time the main shock reaches the edge of the progenitor and starts propagating through the shell material. By the time when the main shock reached the shell's edge, the shell expanded to another 100--200~\Rsun{} and represented a very compact CSM, i.e. the shell is unlikely to have a significant effect on the resulting LC, as we showed in Section~\ref{subsect:lcs} for models W12--W17 with a compact CSM. The shell matter is ionised and optically thick, therefore the radiation from the front of the main shock is not seen until it breaks out from the shell and continues moving through the lower-density CSM. The LC is formed as a combination of the flux from the shell ejection with its energy of $0.01-0.02$~foe and from the SN shock that was initiated by the release of 0.7~foe in the explosion mechanism.

We emphasise that aspherical mass ejection by 3D asymmetries of the SN explosion as well as any deviation of the progenitor from ideal sphericity, caused by large-scale convective plumes or the outward propagation of gravity waves, may play an important role in shaping the overall LC. The visible SN emission includes flux contributions from different components of the ejecta, thus probing larger volumes and deeper layers as time progresses.

\subsection[SN precursors as probes of new physics]{SN precursors as probes of new physics}
\label{subsect:exotic}

In this section, we highlight that early-time observations of SN explosions, such as those of SN\,2023ixf analysed in this work, might offer a promising avenue for exploring physics beyond the Standard Model. Conclusions on such new physics, however, are reasonable and evidential only if the effects of asymmetries in the mass distribution of the progenitor envelope, explosion ejecta, and of the matter in the closest vicinity of the pre-explosion star are well understood.

SNe, particularly SN\,1987A (due to its registered neutrino signal), have been routinely used to impose some of the strongest constraints on new particles, such as axions, dark photons, sterile neutrinos, etc. \citep{Raffelt:1996wa, 2024cacw.confE..41C}. The underlying principle is that, even though exotic particles interact only weakly with Standard Model particles, they can still be produced in large numbers within the dense and hot cores of SNe, much like neutrinos. Depending on their masses and couplings strengths, these particles can give rise to distinctive observational signatures, potentially leading to their discovery or imposing limits on their existence. To illustrate this idea, we focus on a specific example: axion-like particles coupled to photons, which are among the most studied hypothetical  particles nowadays \citep{Mirizzi:2016eza, DiLuzio:2020wdo, 2024cacw.confE..41C, OHare:2024nmr, Safdi:2022xkm}. While here we use axions\footnote{In this work, we use the terms ``axion'' and ``axion-like particles'' interchangeably. More precisely, particle physicists typically use ``axions'' to refer to pseudoscalar bosons that solve the strong Charge-Parity (CP) problem \citep{Peccei:1977hh, Weinberg:1977ma, Wilczek:1977pj} and primarily couple to nucleons, though they can also couple to photons. In this case, the particle's mass and its coupling to photons are related. In contrast, ``axion-like particles'' are pseudoscalar bosons that do not necessarily solve the strong CP problem, and for which mass and photon coupling are independent parameters.} as a jumping-off point, the general framework applies, with appropriate modifications, to a wide range of other exotic candidates.

Consider an axion particle with mass $m_a$ and a coupling to photons $G_{a\gamma\gamma}$. This latter sets the strength of the interaction of axions and ordinary photons, via the interaction term $\mathcal{L}_a \supset - G_{a\gamma \gamma} a \, \vec{E}\cdot \vec{B}$, where $a$ is the axion field and $\vec{E}, \vec{B}$ are the electric and magnetic fields, respectively. Thanks to this interaction term, axions get produced in the SN core either via the Primakoff effect, where photons get converted into axions in the electric field of ambient ions/protons, $\gamma + Ze \rightarrow Ze + a$, or via photon coalescence, $\gamma + \gamma \rightarrow a$. If the axion mass is above $\sim 50 \, \rm MeV$ or so, coalescence is the most important channel, otherwise the Primakoff effect dominates. 

After being produced in the inner part of the star, axions travel outwards relativistically (or semi-relativistically, if the mass is very large) and their fate crucially depends on their mean free path to decay back into a pair of photons, $\lambda_{a \rightarrow \gamma\gamma}$. This mean free path is similar to the radiative decay of pions and reads
\begin{equation}
\lambda_{a \rightarrow \gamma\gamma} = \frac{64 \pi}{G_{a\gamma\gamma}^2}\frac{\sqrt{\omega_a^2-m_a^2}}{m_a^4} \simeq \frac{4 \times 10^{13} }{G_9^2}\frac{\omega_{\rm 100}}{m_{10}^4}\, \rm cm,
\end{equation}
where $G_9 \equiv G_{a\gamma\gamma}/10^{-9} \rm GeV^{-1}$, $\omega_{100}\equiv \omega_a/100\,\rm MeV$ is the particle energy of the axions, $m_{10} \equiv m_a/ 10\,\rm MeV$, and where in the last part of the equation we have taken the relativistic limit $\omega_a \gg m_a$ for simplicity.

\begin{figure} 
\centering 
\hspace{-5mm}\includegraphics[width=0.5\textwidth]{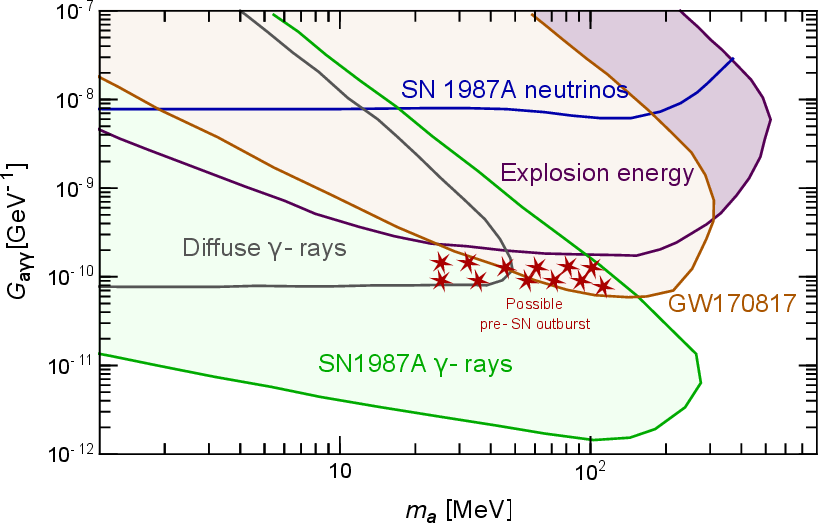} 
\caption{Axion parameter space. All shaded regions are excluded by one or more astrophysical probes. Specifically, the region above the blue line is excluded by the duration of the neutrino signal from SN\,1987A ~\citep{Raffelt:1996wa}. The purple region is excluded by the explosion energies of Type IIP SNe ~\citep{2022PhRvL.128v1103C}, assuming a typical star with a radius of $R_* = 5 \times 10^{13} \rm cm$ and a lower limit on the explosion energy of 0.1~foe. The gray region is constrained by measurements of the extragalactic background light~\citep{Caputo:2021rux}, while the green region is excluded by the upper limit on the gamma-ray fluence from SN\,1987A~\citep{Caputo:2021rux}, as determined by GRS. The shaded orange region is instead in tension with X-ray observations of GW170817 event~\citep{Diamond:2023cto}. The red stars highlight points of the parameter space around which the axion's mean free path is within $\sim$$10^{\rm 12}-10^{\rm 14} \, \rm cm$ and the energy deposition at such distances accrues to $\sim$$0.01-0.1$\,foe.} 
\label{figure:axions} 
\end{figure}

Thus, if axions are very light and/or have a very small coupling, they can freely escape from the star, $\lambda_{a \rightarrow \gamma\gamma} \gg R_*$ (where $R_*$ is stellar radius), effectively providing a new cooling channel. The presence of an extra cooling channel would make the SN core cooling down faster, in turn shortening any associated neutrino signal. This is in essence the ``cooling argument'' that was applied to SN\,1987A almost 50~years ago~\citep{Turner98} (although the first papers on the subject studied the axion coupling to nucleons, rather than photons). As a rough estimate, if the axion luminosity $L_a$ for a given point in parameter space -- defined by mass and coupling -- is comparable to that of neutrinos around one second after core bounce, then that point in the parameter space is excluded. This is the so-called ``Raffelt criterion'' \citep{Raffelt:1996wa}. This limit corresponds to the solid blue line in Figure~\ref{figure:axions}, obtained using the simulations of \citet{Bollig:2020xdr}.

Once axions escape the star, they can still decay into photons outside the progenitor envelope, producing gamma-ray signals. For example, if the mean free path is comparable or smaller than the distance to SN\,1987A, $d_{\rm 1987A} \simeq 51.4 \, \text{kpc}$, strong constraints derive from the Gamma-Ray Spectrometer (GRS) on board the Solar Maximum Mission (SMM) satellite that operated 02/1980–12/1989. Photons from axions emitted by SN 1987A would have been picked up by this instrument, which instead set only upper limits on the photon fluence, thereby constraining the existence of axions~\citep{Jaeckel:2017tud, Caputo:2021rux, Hoof:2022xbe}, as indicated by the green region in Figure~\ref{figure:axions}. Interestingly, \citet{Muller:2023pip} applied the same argument to SN\,2023ixf, using Fermi-LAT gamma-ray observations of the event to put constraints on smaller axion masses (given that SN\,2023ixf took place way more distant than SN\,1987A).

If instead $\lambda_{a \rightarrow \gamma\gamma}$ is of cosmological size, then the main constraints come from the integrated signals of all past SNe. In fact, axions would create a cosmic background density analogous to the diffuse SN neutrino background~\citep{Ando:2004hc, Beacom:2010kk, Mirizzi:2016eza, 2021ApJ...909..169K}; then, when they decay, they contribute to the diffuse gamma-ray background. One can then use the extragalactic background light measured by a variety of missions~\citep{Fermi-LAT:2014ryh}, such as Fermi-LAT or SMM, to place complementary constraints on the axion parameter space~\citep{Caputo:2021rux}, indicated in gray in Figure~\ref{figure:axions}.

So far we have described different observables which can constrain axions with $\lambda_{a \rightarrow \gamma\gamma} \gg R_*$. What if instead $\lambda_{a \rightarrow \gamma\gamma} < R_*$? 
In this case, if $R_{\rm NS} < \lambda_{a \rightarrow \gamma\gamma} < R_*$ (where $R_{\rm NS}$ is the radius of the new-born neutron star), most of the electromagnetic energy is dumped within the progenitor star, contributing to the SN explosion energy, which thus provides a ``calorimetric'' constraint on new particles \citep{Falk:1978kf}. Such a constraint is particularly severe when applied to a SN population with particularly low explosion energies~\citep{Spiro_2014, 2015ApJ...806..225P, 2019ApJ...879....3G}, as recently suggested by \citet{2022PhRvL.128v1103C}. When axions decay on route to energetic gamma-ray photons in the stellar interior, the photons get quickly absorbed by pair production on nuclei. The precise impact of this energy deposition is subtle: if the luminosity in axions is locally comparable or larger than the Eddington luminosity, then the energy deposited will directly accelerate the medium, otherwise it will mostly heat the stellar matter. In either case, the energy deposited must be below approximately $0.1 \, \text{foe}$ to remain consistent with observations~\citep{2022PhRvL.128v1103C}. This bound is shown as a purple shaded region in Figure~\ref{figure:axions}. The lower part of this curve up to masses of $\sim 200 \, \text{MeV}$ is determined by the condition 
$\lambda_{a\rightarrow\gamma\gamma} \sim R_{\rm IIP}$, where $R_{\rm IIP} = 5 \times 10^{\,13} \, \rm cm$ is the typical size of a progenitor of a SN\,IIP  adopted in \citet{2022PhRvL.128v1103C}. Axions with parameters below this curve, and with $m_a \lesssim 200 \, \rm MeV$, have a mean free path longer than $5 \times 10^{\,13} \, \rm cm$ and deposit an energy $0.1 \, \text{foe} \times (G_{a\gamma\gamma}/G_{a\gamma\gamma}^{\rm curve})^2$, where $G_{a\gamma\gamma}^{\rm curve}$ is the axion-photon coupling value at a given mass on the purple curve in Figure~\ref{figure:axions}. Furthermore, in a similar region of the axion parameter space (for slightly larger mean free path), it is also possible that the photons produced from axion decay form a fireball, which would then produce X-ray emission.  \citet{Diamond:2023cto} used X-ray observations of the GW170817/GRB\,170817A event and excluded the region inside the orange curve in Figure~\ref{figure:axions}. Part of the region denoted by ``SN1987A $\gamma-\text{rays}$" is actually excluded by an analogous argument about the production of X-ray, which should have been detected by the Pioneer Venus Orbiter~\citep{Diamond:2023scc}. Actually, one could use Chandra-ACIS and NuSTAR X-ray observations of SN 2023ixf~\citep{2024ApJ...963L...4C, 2023ApJ...952L...3G} in a similar way. 

It is important to notice that the gravitational binding energy of all layers outside the stellar core is at most some $0.01 \, \rm \text{foe}$ in low-mass progenitors (see Table~1 of \citealt{2020MNRAS.496.2039S}), and orders of magnitude lower in the hydrogen envelope. Therefore, the small amount of energy deposited by axions can easily eject the stellar material. As an example, consider the red stars in Figure~\ref{figure:axions}. They identify axions with mass and coupling such that $\lambda_{a\rightarrow\gamma\gamma} \sim 10^{\,12}-10^{\,14} \, \rm cm$, and would deposit an energy of roughly $\gtrsim 0.01$ foe, when decaying. In summary, these axions would: get produced mainly via photon coalescence in the SN core, around 
$r_0 \sim 10 \, \rm km$; travel through the stellar envelope at the speed of light and deposit most of their energy at $r_1 \sim \lambda_{a\gamma\gamma}$, in a time 
$\lambda_{a\gamma\gamma}/c \simeq 1 \, \text{h} \Big( \lambda_{a\gamma\gamma}\, \Big/ 10^{14}\rm cm\Big)$. On the other hand, the main shock forms close to the neutron star as neutrinos interact with the ambient matter, and it propagates to the surface of the star during a time 
$t_\mathrm{sbo}=R_* / v_\mathrm{sh} \sim 0.6\Big( R_*\, \Big/ 5 \times 10^{\,13}\rm cm \Big) \Big( 10^{\,9} \text{cm}\,\text{s}^{\,-1}\, \Big/ v_\mathrm{sh} \Big)\rm day$ (approximate time between the formation of the shock and its breakout from  the stellar surface), where $v_\mathrm{sh}$ is the shock velocity. This implies a time gap of roughly half a day between the energy deposition by axions and the shock breakout. For the parameter space around the red stars in Figure~\ref{figure:axions}, and more generally for parameters lying between the purple region and the green one, the axion energy would get deposited into the external shells studied in this work, at $700-1000 \, \rm R_\odot$. Even if the mean free path is larger than this (which means, for a fixed axion mass, smaller axion-photon couplings), the phenomenology may still be very interesting, with energy injected directly into the outer parts of the CSM around the progenitor star. The red stars in Fig.~\ref{figure:axions} are thus only a rough indication of the parameters for which we expect interesting features to arise in the SN LCs. Notice that the progenitor of SN~1987A was a blue supergiant star with a radius of $(2-3)\times 10^{12}\, \rm cm$~\citep{Woosley:1988at}. Therefore, for the mass and coupling of axions decaying inside the progenitor of SN~2023ixf, the axion decay would have happened outside the progenitor of SN~1987A, producing a detectable gamma-ray signal. 

While the interesting region of the axion parameter space appears to conflict with bounds from GW170817 and SN1987A $\gamma$-ray observations, these two probes are based on single events (unlike the constraints from low-energy SNe or diffuse $\gamma$-ray observations). Since all astrophysical systems have inherent uncertainties --\,which can certainly affect bounds at a factor of a few level\,-- it is always preferable to have multiple probes covering the same region of parameter space.

The interaction of SN ejecta with the external layers of the star and the dense CSM can also produce high-energy cosmic rays, neutrinos and gamma-rays, a possibility recently studied by~\citet{2024arXiv240918935K}. In fact, similar processes may take place also after the energy injection due to axions. Moreover, while in this section we discussed only the phenomenology of massive axions, which can decay on route into gamma-rays, another possibility is that much lighter particles can convert into gamma-rays in the magnetic fields of the progenitor stars~\citep{2024PhRvL.133u1002M}, if these fields are strong enough.

In summary, observations of SNe during their early hours can probe interesting regions for axions and other exotic particles. With advancements in transient observation facilities, a new window for investigating novel physics may soon open, encompassing an entire class of events rather than relying on single ones. However, as noted at the beginning of this section, properly characterising the injected energy is intricate, and all these possibilities deserve further detailed studies, which we leave for future work.

\subsection[SN precursor by an aspherical shock breakout]{SN precursor by an aspherical shock breakout}
\label{subsect:3DSN}

Yet another explanation of the early steep rise of the LC of SN\,2023ixf may be offered by the possibility that the explosion itself was intrinsically highly aspherical. One of the examples of an asymmetric explosion is the low-mass (9\,\Msun{}) model s9.0 that was evolved self-consistently in 3D from the onset of the neutrino-driven explosion until shock breakout \citep{2020MNRAS.496.2039S}. In such simulations the SN shock gets deformed during the first seconds after core bounce because neutrino heating triggers violent convective activity and buoyant mass flows in the postshock volume. These initial ejecta asymmetries seed the growth of secondary Rayleigh-Taylor instability when the outgoing SN shock accelerates and decelerates around composition interfaces in the progenitor. The Rayleigh-Taylor plumes and associated Kelvin-Helmholtz instability lead to efficient mixing, by which heavy elements are carried outward and helium and hydrogen inward. Moreover, a large-scale asymmetry is imposed on the ejecta, which in extreme cases and under favorable conditions \citep[as in the s9.0 model of][]{2020MNRAS.496.2039S} can even produce a global deformation of the SN shock when propagating through the stellar envelope.

The deformed shock implies different velocities and energies in different directions and therefore it reaches the stellar surface at different times. The shock breakout signal appears first in the direction of the fastest Rayleigh-Taylor plumes and therefore the time between the shock breakout of the fastest plume and the slowest (or the average) parts of the SN shock is up to one day in model s9.0. Such intrinsic explosion asymmetries might concur with large-scale convective asymmetries in the RSG envelope to stretch the breakout emission from the deformed shock over even longer times \citep{2022ApJ...933..164G} and to produce precursor features \citep[as searched for in SN\,2024ggi;][]{Shrestha+2024}. The case of a 9\,\Msun{} explosion model investigated by \citet{2020MNRAS.496.2039S} might be special and future studies by 3D long-term neutrino-driven core-collapse simulations will have to show whether such effects can play a role for a wider variety of CCSNe.
 
\citet{2022MNRAS.514.4173K} evaluated the s9.0 model for its direction dependent LCs (though only approximately by considering 1D transport with the \verb|STELLA| code for different radial slices) and the corresponding $4\pi$-equivalent explosion energies and showed that the breakout behavior and inferred isotropic-equivalent explosion properties can vary significantly for different viewing angles. Since the observable emission depends on the observer position when the actual source is extremely non-spherical, our parametric 1D shell models might imply a misjudgment, i.e., overestimation, of the involved masses and energies.

Effects associated with asymmetric shock breakout can also be expected when the SN shock propagates through a rotationally deformed progenitor or magnetic field amplification in collapsing rotating stars causes magnetorotationally driven jets, jet-like collimated outflows, or choked jets that deposit their energy in expanding cocoons \citep[e.g.,][]{1976Ap&SS..41..287B,2019ApJ...871L..25P,2021MNRAS.503.4942O,2021MNRAS.507..443B,2023MNRAS.522.6070P}. Although current stellar evolution and CCSN theory suggest that these phenomena are probably connected to special or extreme progenitor conditions (e.g., low metallicity, high angular momentum, strong magnetic fields) they might play a role for a wider spectrum of CC events \citep{2019ApJ...871L..25P,Soker2024} and deserve further exploration.  


\section[Conclusions]{Conclusions}
\label{sect:conclusions}

In this study, we built a progenitor model for SN\,IIP 2023ixf, whose explosion with an energy of 0.7~foe is in agreement with the explosion energies of average Type~IIP SNe according to observational and theoretical estimates. Our progenitor is based on a 10~\Msun{} stellar evolution model; however, we do not exclude that it could also be a massive star with a higher initial mass. We assumed that the progenitor has lost quite a large fraction of its hydrogen-rich envelope, more than it could strip via mass-loss in a steady wind. We suggest that the higher mass-loss was caused by binary interaction. We explain the peak of SN\,2023ixf by the presence of a wind-like CSM, preferentially with a radius of $10^{15}$~cm and a mass of $0.3-0.83$~\Msun{}. This CSM might have formed as a result of stronger, late mass-loss caused by gravity waves driven by convective nuclear burning in the inner shells during the final years of the star's evolution.

However, the early steep rise in the LC of SN\,2023ixf, which is seen during the first five hours after observationally confirmed non-detection and prior to the SN discovery announced by \citet{2023TNSTR..39....1I}, cannot be reproduced in our 1D simulations by CSM interaction of the SN shock in any of our otherwise well fitting models. We therefore hypothesise that this earliest LC feature might represent precursor emission that could be connected to a pre-SN outburst in the outermost layers (0.5--1\,M$_\odot$) of the progenitor, driven by the deposition of a relatively small amount of energy of $0.01-0.02$~foe near the surface of the hydrogen-rich envelope. Most likely, the particular outburst-causing event did not form the CSM needed to explain the peak of SN\,2023ixf, because there was no time for the ejected shell to reach the radius of $10^{15}$~cm.  
Within the established theoretical possibilities, the source of the energy injection into the hydrogen-rich envelope could be multi-dimensional phenomena such as hydrodynamic instabilities in the shell-burning layers, gravity waves, pulsations, or mass ejection caused by large-scale convective plumes in the stellar envelope, which, in turn, could amplify the RSG wind, However, for all these phenomena an energy of $0.01-0.02$~foe within a short period of time appears to be on the extreme side and hard to reach. 

If the LC precursor is connected to the SN explosion itself, it may point to a highly asymmetric explosion. One possibility is that the SN shock was strongly deformed, either because of a non-spherical progenitor due to rapid rotation or 3D hydrodynamic activity, or because of convective postshock plumes that deformed the shock and thus made it expand with significantly different velocities in different directions. Alternatively, the highly asymmetric explosion could have been caused by a low-energy jet that propagated (maybe even close to the speed of light) well ahead of the SN shock. Any such asymmetric explosion would have an impact on the early rise of the LC.

Sticking to the 1D setups considered in our present paper, yet another possibility is that an energy of $0.01-0.02$~foe was released in the outermost regions of the envelope of the progenitor by the decay of exotic particles, e.g., axions or axion-like particles. Hypothetically, such particles can be produced simultaneously with neutrinos during the collapse of the stellar core and the cooling of the newly formed, hot neutron star. However, this hypothesis requires a separate thorough study and will be conducted in the future.

The kind of precursor emission analysed in the present study can be discovered during the first hours after the last observationally confirmed non-detection of the SN, and sensitive instruments\footnote{Note that the first data published in \citet{2024Natur.627..754L} were obtained by amateur astronomers with their small telescopes, such as 8-cm, 10-cm, 10.1-cm, 13-cm, 15-cm, 20-cm refractors \citep[see details in][]{2023TNSAN.130....1M}. Certainly, this was a fortunate success because of the unique proximity of SN\,2023ixf.} and advanced software are needed to identify this signal and to extract it from the rapidly swelling overall flux of radiation. Therefore, very high-cadence observations --\,up to minutes\,-- are required to fulfill this task. Operating and upcoming facilities dedicated to observing transients such as LSST, or a wide network of small and medium telescopes \citep{2023TNSAN.213....1B,2023RNAAS...7..141S}, will be able to capture SNe at their birth, enhancing our potential to spot their precursor emission \citep{2024arXiv240813314G}. Such discoveries will help to clarify what is happening at the surface of massive stars while their cores collapse. This is especially important for any next close SN and ultimately for a SN in our own Galaxy. It will open a unique window to test and advance stellar astrophysics and to look for effects beyond the Standard Model of particle physics.

\begin{acknowledgements}
We thank Kirill Sokolovskiy, Pavel Abolmasov, Steve Schulze, Nikolay Pavlyuk, Dmitriy Tsvetkov, Sergei Blinnikov, Edoardo Vitagliano for helpful discussions and comments, Avinash Singh, Brian Hsu, and Xiaofeng Wang for providing data of SN\,2023ixf. 
PB is supported by the grant  RSF\,24-12-00141 for modeling supernova light curves with the \STELLA\, code. AC is supported by an ERC STG grant (``AstroDarkLS'', grant No. 101117510).
HTJ acknowledges support by the German Research Foundation (DFG) through the Collaborative Research Centre ``Neutrinos and Dark Matter in Astro- and Particle Physics (NDM),'' grant No. SFB-1258-283604770, and under Germany's Excellence Strategy through the Cluster of Excellence ORIGINS EXC-2094-390783311.

\end{acknowledgements}


\bibliographystyle{aa}


\begin{appendix}

\onecolumn

\section[Modifications made to the original progenitor model s10]{Modifications made to the original progenitor model s10 \citep{2016ApJ...821...38S}}
\label{appendix:append1}

\begin{figure}[!h]
\centering
\hspace{-5mm}\includegraphics[width=0.49\textwidth]{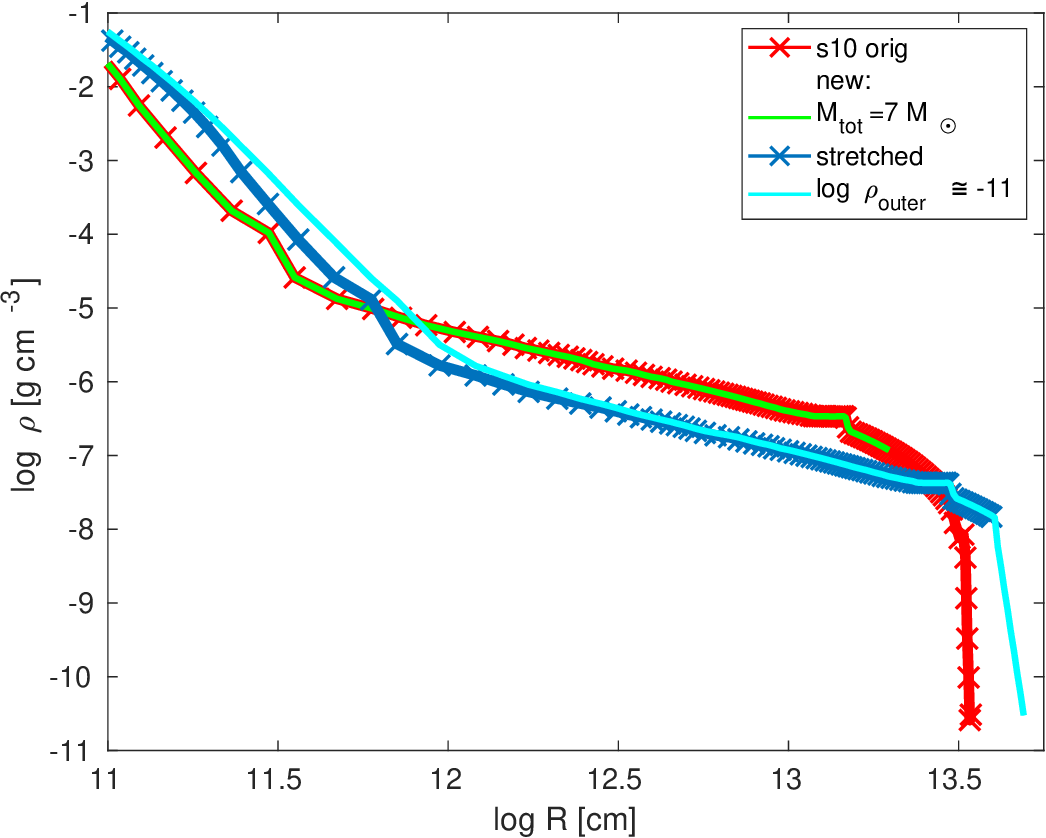}
\caption{Density distributions for the modified progenitor structure (with outer layers cut, then stretched, and with an exponentially declining layer attached) versus radius, compared to the original s10 pre-collapse model. See text for details.}
\label{figure:modiR}
\end{figure}

\begin{figure}[!h]
\centering
\hspace{-5mm}\includegraphics[width=0.49\textwidth]{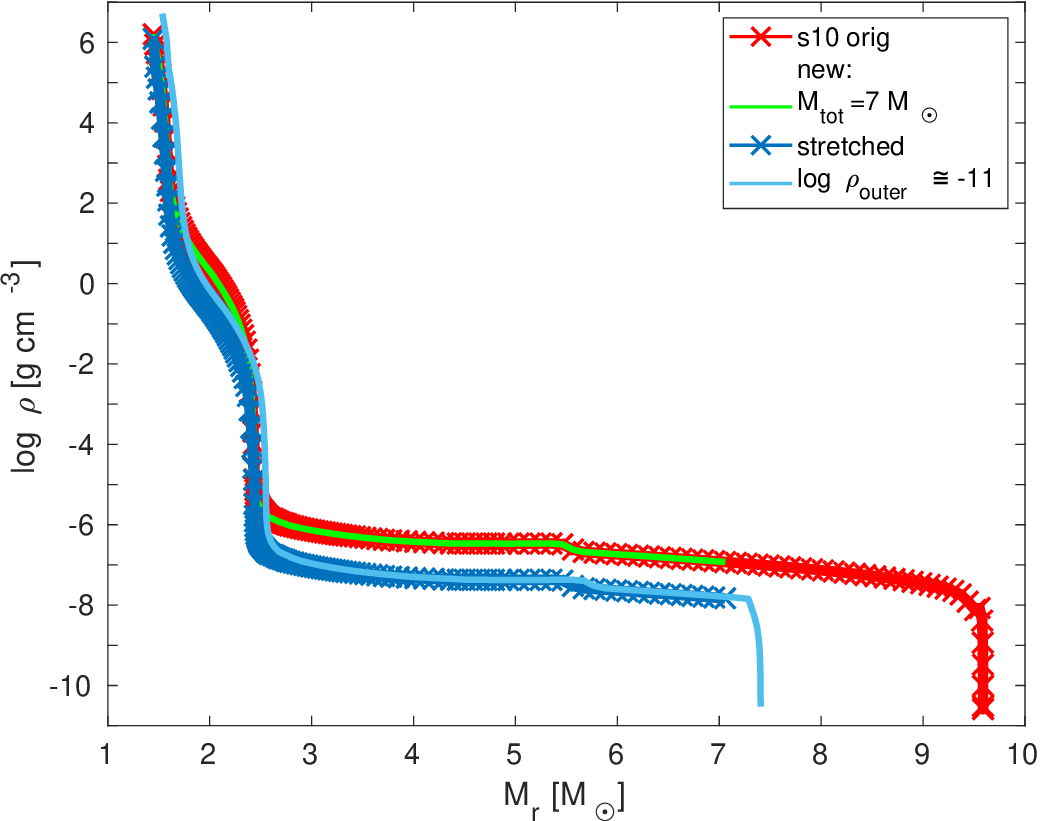}
\caption{Density distribution for the modified progenitor structure (with outer layers cut, then stretched, and with an exponentially declining layer attached) versus mass coordinate, compared to the original s10 pre-collapse model. See text for details.}
\label{figure:modiM}
\end{figure}

Here we explain in more detail the modifications we applied to the original progenitor model from \citet{2016ApJ...821...38S}. This default progenitor model (s10) served as one of five initial models for the large set of LC simulations by \citet{2023PASJ...75..634M}. Moreover, it was found to be a good progenitor for SN~2023ixf in \citet{2024PASJ...76.1050M}. In Figure~\ref{figure:modiR} and Figure~\ref{figure:modiM}, the red curve with times symbols stands for the original progenitor model s10.

\begin{enumerate}
\item We cut the outer layers of the stellar envelope up to the mass
coordinate of 7~\Msun{}. The truncated profile is represented as a green solid line labeled ``$M_\mathrm{tot}=7$~\Msun{}''.

\item We stretched the star while scaling the
density profile and keeping the total mass of the system according to the following relations:
\begin{equation}
\begin{array}{l}
R_\mathrm{new} = R_\mathrm{inner} + (R_\mathrm{old}-R_\mathrm{inner}) \times {\cal R}_\mathrm{factor}\,, \\
\rho_\mathrm{new} = \rho_\mathrm{old} \left[ \frac{\Delta V_\mathrm{old}}{\Delta V_\mathrm{new}}\right] \simeq \rho_\mathrm{old} \left[ \frac{R_\mathrm{old}}{R_\mathrm{new}}\right]^{\,3} \simeq \rho_\mathrm{old} \, {\cal R}_\mathrm{factor}^{\,3}\,,
\end{array}
\end{equation}
where $R_\mathrm{inner}$ corresponds to the inner region, which divides the model into the collapsed stellar core and the ejected stellar matter, and $R$ is the outer radius of a Lagrangian mass zone. $\Delta V_\mathrm{old}$ and $\Delta V_\mathrm{new}$ is the volume of the Langrangian mass zone before and after the stretching procedure. ${\cal R}_\mathrm{factor}$ is a stretching factor, and equals 2 in our modification routine. The relative change in the density value corresponds to $2^{\,3}=8$, i.e. $\log \rho \sim \log 8 \sim 0.9$, which can be seen in Figure~\ref{figure:modiR} and Figure~\ref{figure:modiM}. We note that the progenitor model s10 has a radius of 280~\Rsun{} after the first step of modifications, i.e. cutting the outer layers. The stretched stellar configuration has a radius of 570~\Rsun{}. The stretched model is shown as a blue line with times symbols (label ``stretched'') in Figure~\ref{figure:modiR} and Figure~\ref{figure:modiM}.

\item We appended an artificially constructed layer to force the density to drop to $10^{\,-11}$~g\,cm$^{\,-3}$ according to the power-law expression $\rho\propto r^{\,-30}$, leading to a surface radius of about 700~\Rsun{}. We chose the exponent arbitrary to some extent, however, the density drop imitates well the true progenitor density structure (see, e.g., Figure~\ref{figure:modiM}). The drop of density to the low value is also important for radiative-transfer simulations with \STELLA. Propagation of the shock through the outermost layers depends on the exact density gradient \citep{2017ApJ...845..103L}, if one considers a bare progenitor. For a progenitor surrounded with CSM, the shock does not break out from the optically thick progenitor surface, but instead it immediately starts its propagation through the CSM and radiates from the front, therefore, the streaming radiation would depend more on the parameters of the CSM. Moreover, the mass in this transition layer is 0.4~\Msun{} or low (depending on the point where the CSM is attached), and most likely does not result in a pronounced feature. However, the detailed study is required to analyse this effect, which we leave out in the present paper. 

\item The inner 1.45~\Msun{} are considered to be a newly formed compact object left behind in the core collapse. In \STELLA\, simulations, it is considered as a optically thick part of the computational domain, which produces a gravitational potential.

\end{enumerate}

At the end of the modifications our model, i.e., the central neutron star
($\sim$1.45\,M$_\odot$) plus the ejecta, had a total mass of 7.4~\Msun{}. 

\FloatBarrier 
\clearpage

\section[Appended CSM]{Appended CSM}
\label{appendix:append2}

\begin{figure}[!h]
\centering
\hspace{-5mm}\includegraphics[width=0.49\textwidth]{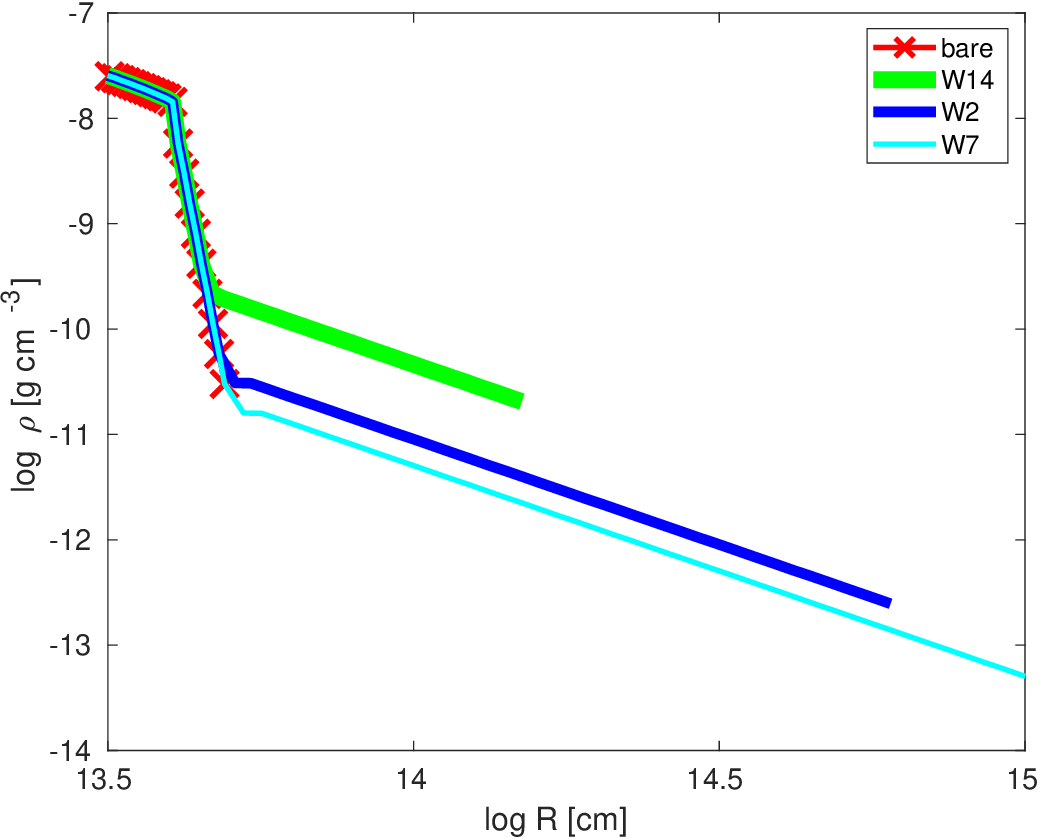}
\caption{Density distribution for wind-models with the CSM mass of 0.3~\Msun{} for three considered CSM radii: $1.5\times10^{\,14}$~cm (W14), $6\times10^{\,14}$~cm (W2), and $10^{\,15}$~cm (W7).}
\label{figure:CSM}
\end{figure}

For the models with CSM, we appended a wind-like layer to the constructed progenitor, i.e. the density distribution is $\rho\propto r^{-2}$ in the CSM. The density at the inner boundary of the CSM spans a range of $(0.3-118)\times10^{\,-11}$~g\,cm$^{\,-3}$ (see Table~\ref{table:models}). 
Figure~\ref{figure:CSM} shows three models with the same CSM mass of 0.3~\Msun{} and three considered CSM radii, $1.5\times10^{14}$~cm, $6\times10^{14}$~cm, and $10^{\,15}$~cm standing for the W14, W2, and W7 models.

\FloatBarrier 
\clearpage

\section[Broad-band LCs]{Broad-band LCs} 
\label{appendix:bands}

\begin{figure*}[!h]
\centering \hspace{-5mm}
\hspace{-1mm}\includegraphics[width=\textwidth]{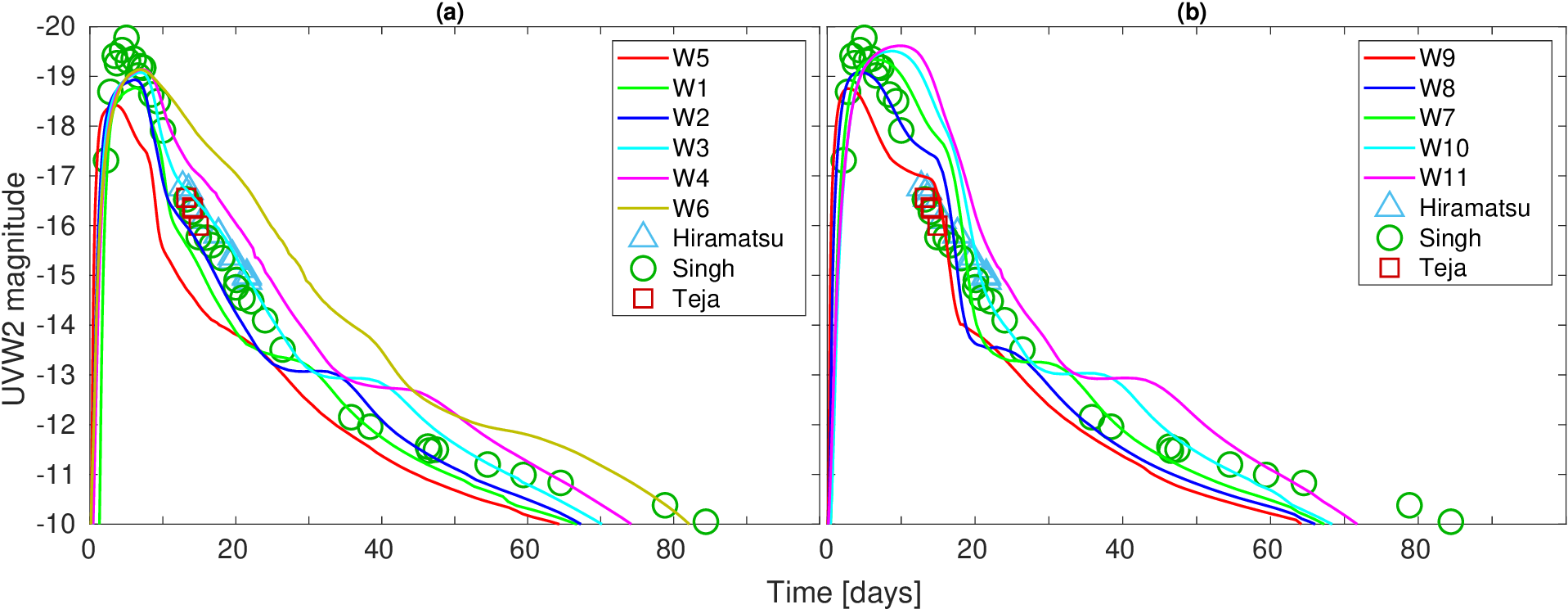} 
\caption{$UVW2$ magnitudes for models W1--W6, $R_\mathrm{CSM}=6\times10^{14}$~cm (a; left) and W7--W11, $R_\mathrm{CSM}=10^{15}$~cm (b; right). The observational data are taken from \citet{2023ApJ...955L...8H}, \citet{2023ApJ...954L..12T}, and \citet{2024ApJ...975..132S}.} 
\label{figure:uvw2_all} 
\end{figure*}

\begin{figure*}[!h]
\centering 
\hspace{-1mm}\includegraphics[width=\textwidth]{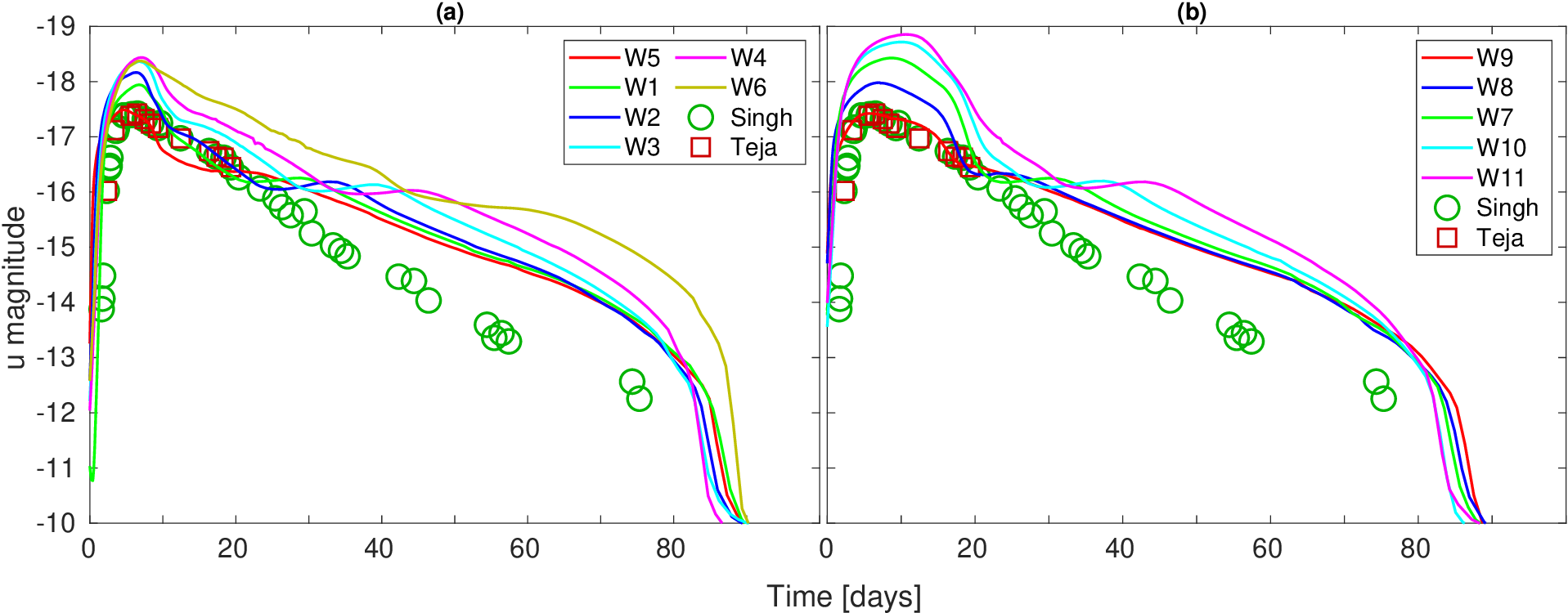}
\caption{$u$ magnitude for models W1--W6, $R_\mathrm{CSM}=6\times10^{14}$~cm (a; left) and W7--W11, $R_\mathrm{CSM}=10^{15}$~cm (b; right). The observational data are taken from \citet{2023ApJ...954L..12T} and \citet{2024ApJ...975..132S}.} 
\label{figure:u_all} 
\end{figure*}

\begin{figure*}[!ht]
\centering 
\hspace{-1mm}\includegraphics[width=\textwidth]{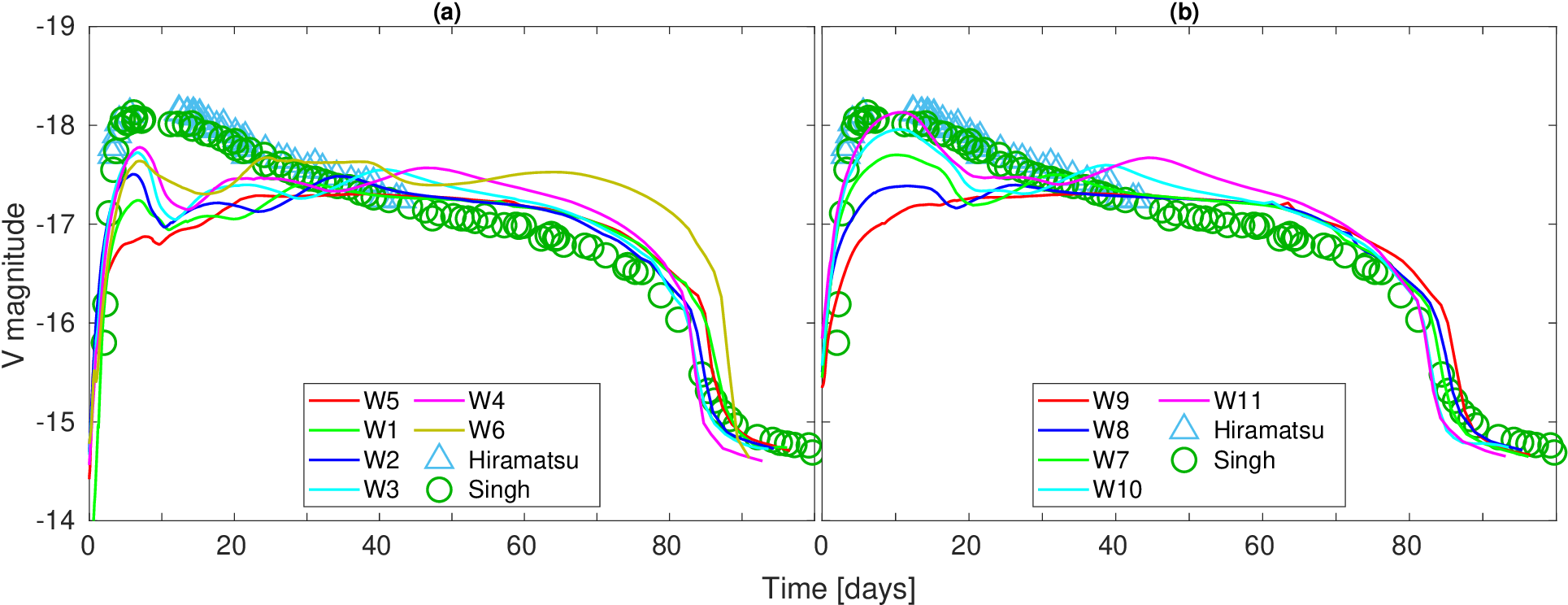}
\caption{$V$ magnitude for models W1--W6, $R_\mathrm{CSM}=6\times10^{14}$~cm (a; left) and W7--W11, $R_\mathrm{CSM}=10^{15}$~cm (b; right). The observational data are taken from \citet{2023ApJ...955L...8H} and \citet{2024ApJ...975..132S}.} 
\label{figure:V_all} 
\end{figure*}

\begin{figure*}[!h]
\centering 
\includegraphics[width=\textwidth]{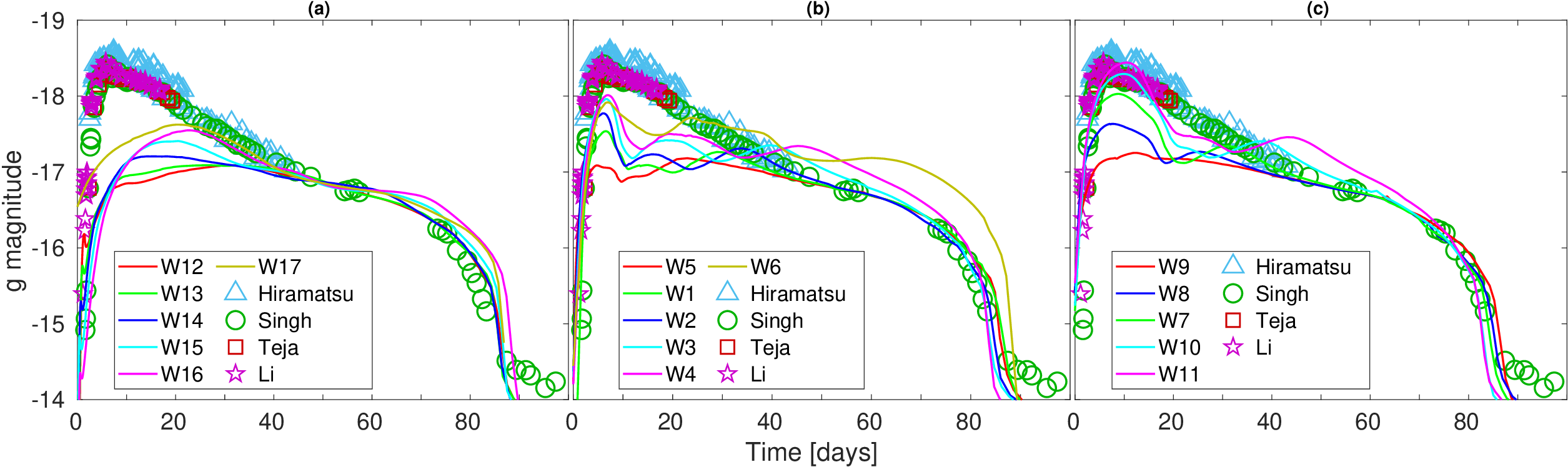}
\caption{$g$ magnitude for models W12--W17, $R_\mathrm{CSM}=1.5\times10^{14}$~cm (a; left); W1--W6, $R_\mathrm{CSM}=6\times10^{14}$~cm (b; middle); and W7--W11, $R_\mathrm{CSM}=10^{15}$~cm (c; right). The observational data are taken from \citet{2023ApJ...955L...8H}, \citet{2023ApJ...954L..12T}, \citet{2024Natur.627..754L}, and \citet{2024ApJ...975..132S}.} 
\label{figure:g_all} 
\end{figure*}

\begin{figure*}[!hb]
\centering 
\includegraphics[width=\textwidth]{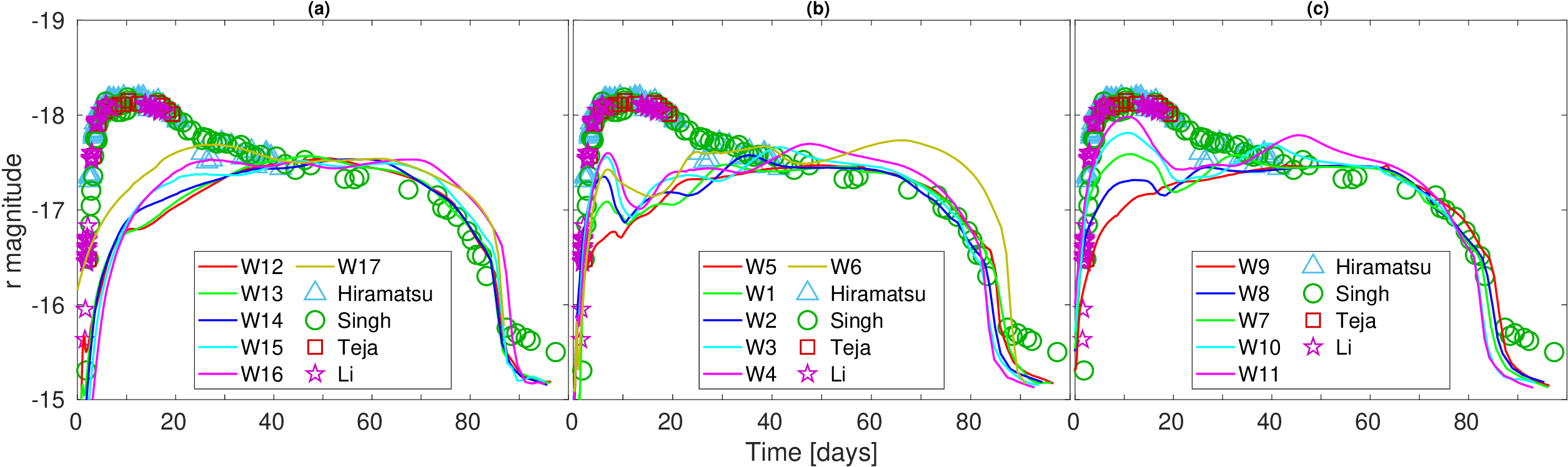}
\caption{$r$ magnitude for models W12--W17, $R_\mathrm{CSM}=1.5\times10^{14}$~cm (a; left); W1--W6, $R_\mathrm{CSM}=6\times10^{14}$~cm (b; middle); and W7--W11, $R_\mathrm{CSM}=10^{15}$~cm (c; right). The observational data are taken from \citet{2023ApJ...955L...8H}, \citet{2023ApJ...954L..12T}, \citet{2024Natur.627..754L}, and \citet{2024ApJ...975..132S}.} 
\label{figure:r_all} 
\end{figure*}

\begin{figure*}[!ht]
\centering 
\hspace{-1mm}\includegraphics[width=\textwidth]{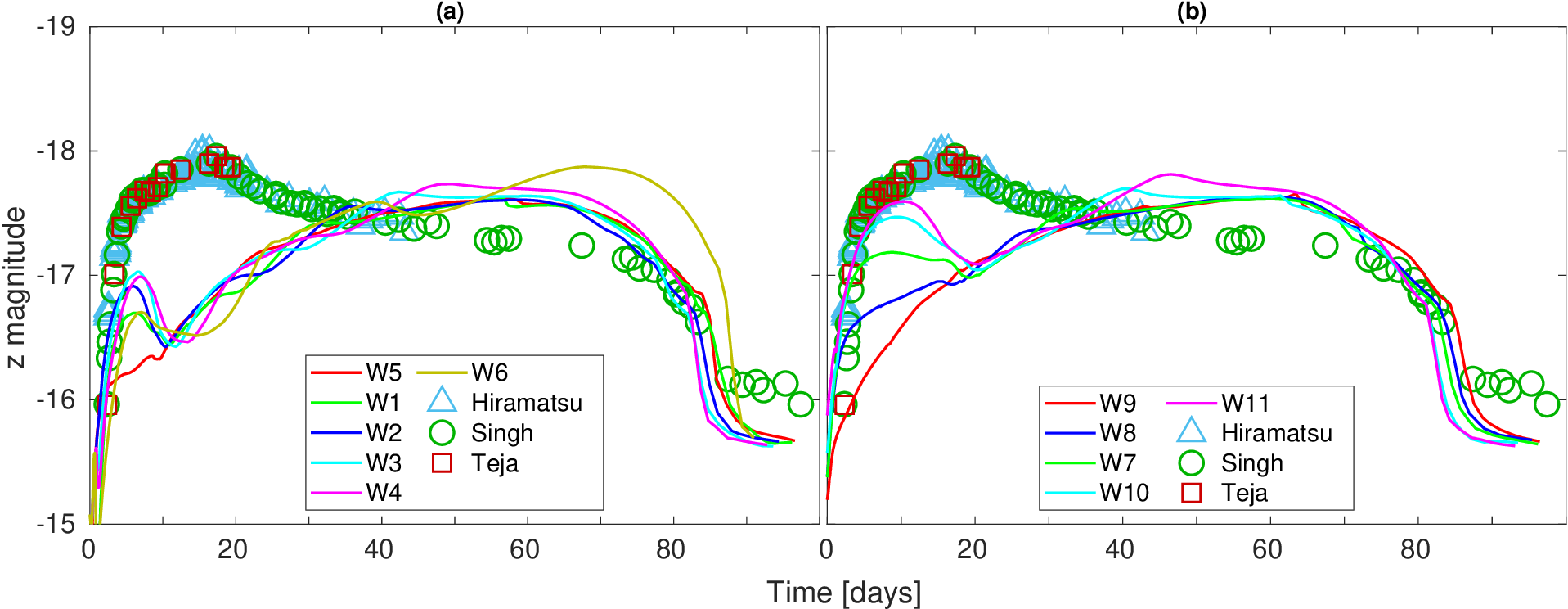}
\caption{{\it z} magnitude for models W1--W6, $R_\mathrm{CSM}=6\times10^{14}$~cm (a; left) and W7--W11, $R_\mathrm{CSM}=10^{15}$~cm (b; right). The observational data are taken from \citet{2023ApJ...955L...8H}, \citet{2023ApJ...954L..12T}, and \citet{2024ApJ...975..132S}.} 
\label{figure:z_all} 
\end{figure*}

\FloatBarrier 
\clearpage

\section[Earliest phase of the SN\,2023ixf LC]{Earliest phase of the SN\,2023ixf LC}
\label{appendix:append3}

\begin{figure}[!h]
\centering
\hspace{-5mm}\includegraphics[width=0.49\textwidth]{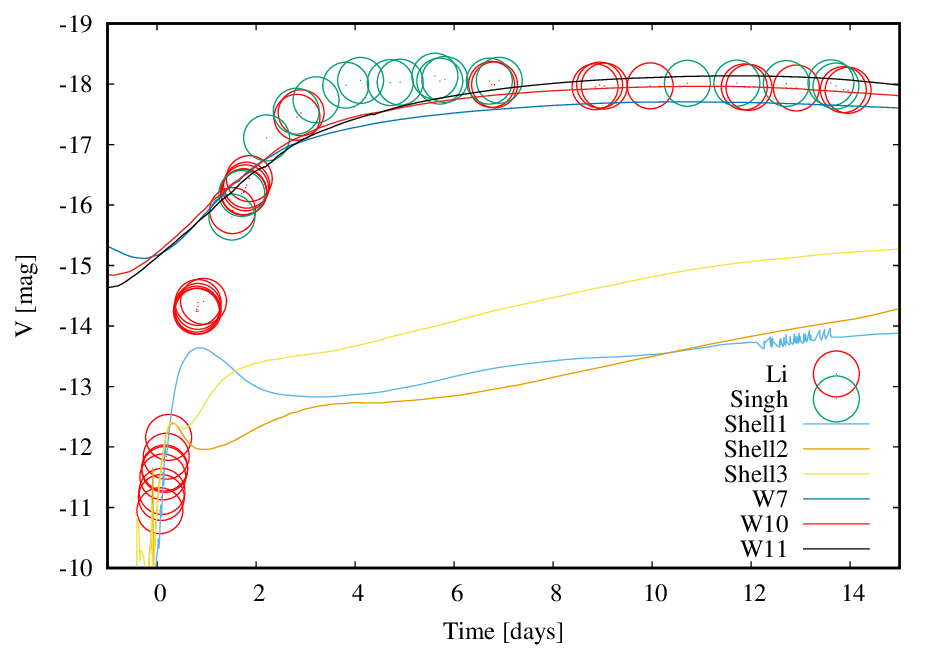}
\caption{$V$-band LCs for our shell-models and the wind-models W7, W10, and W11. The observational data are taken from \citet{2024Natur.627..754L} (``Li'') and \citet{2024ApJ...975..132S} (``Singh'').}
\label{figure:early}
\end{figure}

Here, we demonstrate that the best-fit wind-models, which were built to match the main SN\,2023ixf LC over 100~days, overestimate the flux for the earliest epoch of the SN.
In Figure~\ref{figure:early}, we show $V$-band LCs for the shell-models and wind-models W7, W10, and W11. Contrary to Figure~\ref{figure:bump}, the time interval is 15~days and the upper limit for magnitude is $-$19~mags (versus 16~hours and $-$13.5~mags). Note that the earliest data fall below the lower limit of Figures~\ref{figure:lbol}--\ref{figure:z_all}. 
It is seen that the wind-models match the rising part of the SN\,2023ixf LC, whereas these models have too high a flux during the first day after the discovery. 

\FloatBarrier 
\clearpage

\end{appendix}

\end{document}